# THE HUMAN AUDITORY SYSTEM AND AUDIO

## Milind N. Kunchur

## TABLE OF CONTENTS

# The Human Auditory System and Audio

MILIND N. KUNCHUR
University of South Carolina, Columbia, SC 29208, U.S.A.
Emails: kunchur@gmail.com, kunchur@mailbox.sc.edu
Homepage: http://boson.physics.sc.edu/~kunchur/

This work reviews the human auditory system, elucidating some of the specialized mechanisms and non-linear pathways along the chain of events between physical sound and its perception. Customary relationships between frequency, time, and phase—such as the uncertainty principle—that hold for linear systems, do not apply straightforwardly to the hearing process. Auditory temporal resolution for certain processes can be a hundredth of the period of the signal, and can extend down to the microseconds time scale. The astonishingly large number of variations that correspond to the neural excitation pattern of 30000 auditory nerve fibers, originating from 3500 inner hair cells, explicates the vast capacity of the auditory system for the resolution of sonic detail. And the ear is sensitive enough to detect a basilar-membrane amplitude at the level of a picometer, or about a hundred times smaller than an atom. This article surveys and provides new insights into some of the impressive capabilities of the human auditory system and explores their relationship to fidelity in reproduced sound.

## 1 INTRODUCTION

Music, for many people, is an essential nutrient of life. Most of its consumption, for reasons of practicality and economy, takes place through electronically reproduced *audio*. Unfortunately, listeners accustomed to live acoustic music usually find the audio version to be woefully unrealistic and inaccurate.

Two of the main challenges[1] in reproducing a convincing illusion of a live performance are: (1) *Spatial*—the three-dimensional placement of instruments along with the positional and directional distribution of sonic reflections and reverberant sound field. (2) *Tonal*—related to the timbre of the instrument/s and the performance-room acoustics. Exact spatial recreation cannot be expected because the details of the underlying psychoacoustics and auditory neurophysiology are different for natural-sound *localization* versus stereo *spatialization*[2] [1]. A priori, there is no reason why tonality cannot be exactly reproduced. Still, most audio systems are a long way from reaching this elusive goal. Partly this is because specifications and considerations used in mainstream audio are often based on an overly simplistic view of the hearing process. Standard specifications such as the frequency response (FR) and time-correlated (e.g., harmonic and intermodulation) distortions do not consistently predict perceived sound quality and can even reverse correlate with it[3].

The realm of sound reproduction referred to as *high-end audio* (which will be abbreviated as HEA) takes a no-holds-barred approach in improving sonic accuracy—reducing every possible distortion[4] (measured or postulated) and employing sighted (i.e., not blind) listening tests to steer incremental design changes that may cumulatively make an audible improvement. The lack of insightful measurements, paucity of formal IRB (Institutional Review Board) approved blind listening tests [2] [3] [4], and seemingly extreme and superfluous measures (e.g., atomic clocks, exotic cables, etc.) shroud HEA in skepticism and disbelief. Because of HEA's rarity, many audio consumers are not aware that a well set up 2-channel stereo system is capable of portraying all three dimensions [5] [6].

The present work provides a biological explanation for these enigmas and suggests new types of measurements and blind tests, which can hopefully be incorporated into future audio-equipment evaluation and development. This article also provides a succinct yet detailed description of the chain of events from sound to perception, which should be of value to readers beyond audio and acoustics—those who simply have an interest in the functioning and intricacies of the human auditory system.

---

[1] Some other potential issues are: errors in analog playback speed affecting note durations and tempo, and low-powered systems not being realistically loud enough.
[2] Localization is the process by which the auditory system determines direction and location of a sound source. The term spatialization (or imaging or sound staging) is used to describe an audio system's ability to portray dimensionality somewhat resembling the natural scene. These processes are expounded below.
[3] E.g., injudicious negative feedback can flatten FR and reduce harmonic distortion at the expense of transient response, hurting the overall perceived quality.
[4] Except where specified, the term distortion will be used in the general sense to mean any alteration in waveform.

**GLOSSARY OF ABBREVIATIONS AND SYMBOLS**

| | |
|---|---|
| A1 | primary-auditory cortex |
| AC | auditory cortex |
| AM | amplitude modulation |
| AN | auditory (cochlear) nerve |
| ANF | auditory nerve fiber |
| AT | activation threshold (of an ANF) |
| AVCN | anterior ventral cochlear nucleus |
| BM | basilar membrane |
| $\chi^2$ | chi-squared value (for statistical assessment) |
| CA | cochlear amplifier/amplification |
| CB | critical band/bandwidth |
| CF | characteristic (or best) frequency |
| $C_m$ | membrane capacitance of a neuron |
| CN | cochlear nucleus |
| DAC | digital-to-analog converter |
| DAS | dorsal acoustic stria |
| dB | decibels |
| dB HL | decibels of hearing loss |
| dB SPL | SPL in decibels at a spatial location |
| DCN | dorsal cochlear nucleus |
| DL | difference limen (same as JND) |
| DNLL | dorsal nucleus of the lateral lemniscus |
| DR | dynamic range |
| DRR | direct-to-reverberant (intensity) ratio |
| DSD | direct-stream digital |
| $\Delta t$ | neuronal integration window |
| | also various temporal parameters/delays |
| E | energy (capacity to do work, in J) |
| ELC | equal-level contour |
| EMP | extended-multiple-pass (listening) |
| EPSP | excitatory postsynaptic potential |
| ERB | equivalent rectangular bandwidth |
| $\phi$ | phase (angle) |
| f | frequency |
| $f_c$ | cutoff frequency of audio component |
| $f_{max}$ | pure-tone upper audiometric limit |
| $f_{min}$ | pure-tone lower audiometric limit |
| $f_s$ | sampling frequency for a digital audio system |
| FM | frequency modulation |
| FR | frequency response |
| FSF | frequency sharpening feedback |
| FWHM | full width half maximum |
| GBC | globular bushy cell |
| $G_{KL}$ | low-threshold K$^+$ (ionic) conductance |
| HEA | high-end audio |
| HG | Hechl's gyrus (cortical region containing A1) |
| HRTF | head related transfer function |
| I | intensity of sound (in W/m$^2$) |
| $I_0$ | standard threshold audible intensity of 1 pW/m$^2$ |
| IC | inferior colliculus/colliculi |
| IE | inhibitory-excitatory |
| IHC | inner hair cell |
| ILD | inter-aural level difference |
| IMD | intermodulation distortion |
| IPSP | inhibitory postsynaptic potential |
| ISO | International Organization for Standardization |
| ITD | inter-aural time difference |
| J | joule (unit of energy and work) |
| JND | just noticeable difference |
| L | Sound intensity level (in dB) |
| LGB | lateral geniculate body (or complex) |
| LL | lateral lemniscus |
| LNTB | lateral nucleus of the trapezoid body |
| LOC | lateral olivocochlear system |
| LSO | lateral superior olive |
| LTD | long-term depression (of synaptic connectivity) |
| LTP | long-term potentiation (of synaptic connectivity) |
| MF | mechanical feedback |
| MGB | medial geniculate body (or complex) |
| MNTB | medial nucleus of the trapezoid body |
| MOC | medial olivocochlear system |
| MSN | medullary somatosensory nuclei |
| MSO | medial superior olive |
| NEP | neural excitation pattern (of ANFs) |
| OC | octopus cell |
| OHC | outer hair cell |
| p | p-value (for statistical assessment) |
| P | power (rate of doing work, in W) |
| PCM | pulse-code modulation |
| PVCN | posterior ventral cochlear nucleus |
| r' | auditory perceived distance |
| RD | resolution of detail |
| $R_{in}$ | input resistance |
| $R_{leak}$ | leak resistance |
| s | second/s |
| SBC | spherical bushy cell |
| SC | superior colliculus/colliculi |
| SFR | spontaneous firing rate |
| SNR | signal-to-noise (power) ratio in dB |
| SGC | spiral ganglion cell |
| SOC | superior olivary complex |
| SPL | sound pressure level (numerically similar to L) |
| SPN | superior paraolivary nucleus |
| SSC | short-segment-comparison (listening) |
| SSF | spatial sharpening feedback |
| $\theta$ | angle or angular separation |
| $\tau$ | (audio) temporal smear/resolution |
| $\tau^*$ | (digital-audio) time-shift discrimination |
| $\tau_{60}$ | 60-dB fall time |
| $\tau_c$ | cutoff time (of decay) |
| $\tau_{cell}$ | time constant of a neuron (nerve cell) |
| t | time |
| T | period of oscillation (= 1/f) |
| TM | tectorial membrane |
| TR | (auditory) transient resolution |
| v | speed of sound in air |
| V | electric voltage or potential |
| VAS | ventral acoustic stria |
| VCN | ventral cochlear nucleus |
| VNLL | ventral nucleus of the lateral lemniscus |
| VNLLv | ventral subdivision of the VNLL |
| *W* | work (in J) |
| W | watt (unit of power) |
| x | distance along basilar membrane from apex |

## 2 PHYSIOLOGY OF THE EAR

### 2.1 External and middle ear

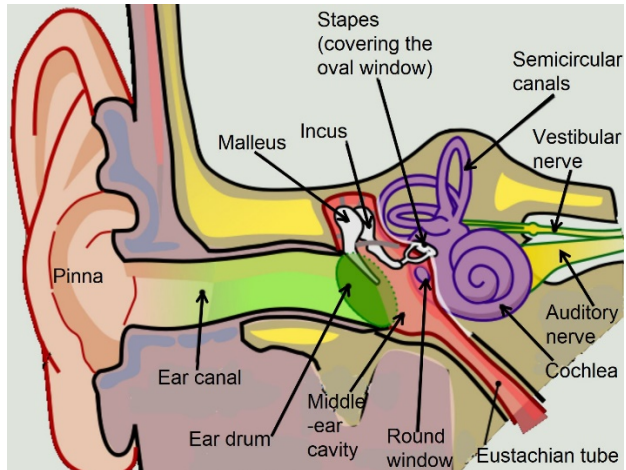

*Fig. 1 Diagram of the human ear (based on [7]). The vestibular system comprised of the semicircular canals is associated with balance, not hearing, but together with the cochlea comprises the 'inner ear'. The cavity between the oval window and eardrum, connected by the eustachian tube to the pharynx, is termed 'middle ear'. The eardrum, ear canal, and pinna comprise the 'external ear'.*

Fig. 1 shows a diagram of the human ear. Sound enters the external ear through the pinna (or auricle), traverses the ear canal (or external auditory meatus), and impinges on the eardrum (or tympanum or tympanic membrane)[5]. The eardrum is attached to a linkage of three miniscule bones in the middle ear—malleus, incus, and stapes (or hammer, anvil, and stirrup) collectively called ossicles[6]. The stapes pushes the vibrations into the cochlea in the inner ear through the oval window. The ossicles, approximating a class-1 lever, amplify the force by 1.3 times. This together with the 20-fold hydraulic gain (due to the 20:1 eardrum to oval-window area ratio) boosts the final pressure by 26 times. This impedance matching is necessary to efficiently couple vibrational energy from air into the cochlea's liquid environment. In its passage to the cochlea, the sound's level[7] is actively adjusted (above ~85 dB) by the protective acoustic reflex mechanism: the tensor tympani muscle acting on the malleus tightens the ear drum, and the stapedius muscle[8] reduces the stapes-to-cochlea coupling. Also the spectrum is resonantly boosted in the region of the speech frequencies in three successive stages. As expounded below, this spectral shaping can be modeled by an inversion of the *equal-level contours* (ELC) and noise data [8].

### 2.2 Cochlea

The *cochlea* (Latin word for snail) consists of a ~35±5 mm long [9] conduit of three parallel *scalae* (or canals or ducts) wound spirally by 2¾ turns into a structure that looks like a snail. A simplified longitudinal section is shown in Fig. 2. The stapes pushing on the oval window (also see Fig. 1) sends a traveling wave through the *scala vestibuli* (or vestibular canal) to its end, where it makes a U-turn through the *helicotrema*, returns through the *scala tympani* (or tympanic canal), and exits the cochlea through the round window back into the middle ear. Wedged between the scala vestibuli and scala tympani lies the *scala media* (or cochlear duct).

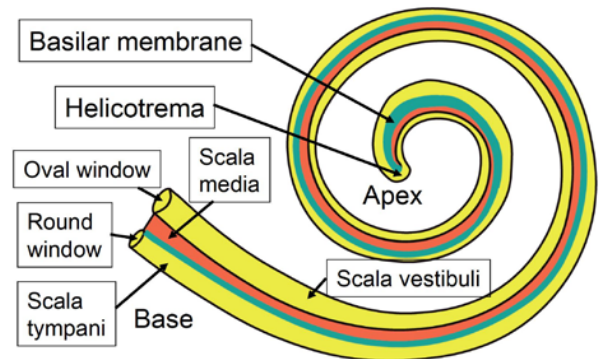

*Fig. 2 Simplified longitudinal section of the cochlear conduit. The end near the middle ear is termed the 'base', and the far end the 'apex'. Scalae tympani and vestibuli contain perilymphatic fluid (yellow), whereas scala media contains endolymphatic fluid (orange) with a higher $K^+$ ion concentration. The basilar membrane (blue) between scalae tympani and media is progressively tapered in width and stiffness across its length, so that the basal end resonates at high frequencies and the apical end at low frequencies.*

Fig. 3 shows a cross-sectional view of the cochlear conduit. Scalae media and tympani are separated by the basilar membrane (BM), in which are embedded ~3500 rows of transducing receptor cells, with one *inner hair cell* (IHC, performing mainly as a "microphone") and 3 or 4 *outer hair cells* (OHCs, performing mainly as "speakers") per row[9]. The cross section of Fig. 3 shows just one row, but the "unfolded" BM of Fig. 4(a) schematizes how rows are arranged over its length. The BM becomes progressively narrower (from ~0.5 to ~0.1 mm) and stiffer going from its apex to base (end near the oval window) [10]. So the *characteristic frequency*[10] (CF), at which a section vibrates maximally, increases logarithmically by an octave per distance increment Δx ~ 4 mm, over the ~9 octaves of CF. This progression, as a function of the fractional distance x from the apex, is approximately modeled by the Greenwood function with the constants A=165.4, α=2.1, and k=0.88 for humans [11], [12]:

$$CF = A [10^{\alpha x} - k] \quad (1)$$

---

[5] To facilitate integrating this work with other writings on this subject, common synonyms are listed in parenthesis.
[6] The stapes is the smallest bone in the human body. Also ossicles mature at birth and do not grow thereafter.
[7] A note on "sound level": *Sound intensity* (in W/m²) $I$ = power/area. *Sound intensity level* $L = 10 \log (I/I_0)$ in dB, where $I_0 = 1$ pW/m² is the nominal threshold of hearing. *Sound pressure level* $SPL = 20 \log (P/P_0)$ in dB, where $P$ is the actual rms pressure variation and $P_0 = 20$ µPa is the threshold rms value. In practice, $L \approx SPL$ and both are simply called "sound level" (at 20° C, $P_0 = [I_0 \rho_a v_s]^{1/2} = 20.3$ µPa is close to the nominal 20 µPa, with the density of air $\rho_a = 1.204$ kg/m³ and the sound speed $v_s = 343$ m/s).
[8] At ~6 mm length, it is the smallest skeletal muscle.
[9] Only mammals have OHCs.
[10] Also referred to as *center frequency* or *best frequency*.

This location dependent tuning is referred to as *tonotopy*.

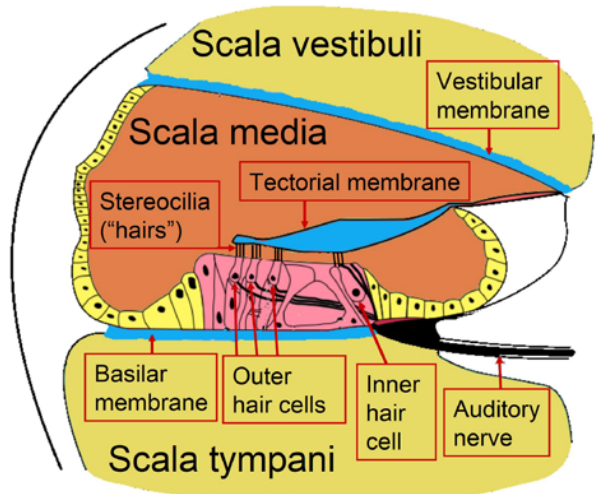

*Fig. 3 Cross section of the cochlear conduit (based on [13]). The 'Organ of Corti', responsible for the transduction of sound, comprises the structure between the basilar and tectorial membranes.*

Relative motion between BM and TM (tectorial membrane), induced by the traveling wave [14], causes IHC *stereocilia* ("hairs") to flex against the TM. The flexing opens mechanoelectrical transduction channels (gates) that admit $K^+$ (potassium) ions to cause a time varying voltage as shown in Fig. 4(b). Then voltage activated gates admit $Ca^{++}$ (calcium) ions, stimulating glutamate neurotransmitter release into synapses with *afferent* (i.e., carrying ascending signals to higher centers) *auditory nerve fibers* (ANFs). The ~8 ANFs per IHC have a range of *activation thresholds* (AT) and *spontaneous firing rates* (SFR) [15], providing ~30000 ANFs labeled by level and frequency. Besides this ANF labeling, level is also encoded in the spike firing rates and frequency is also temporally encoded in the firing pattern. The *neural excitation pattern*[11] (NEP) of the ANFs represents the cochlear information output.

There is a certain amount of cross coupling between different BM regions through the embedding liquid environment, as the wave speed in the liquid (~1 km/s) is much higher than the average propagation speed along the BM (~22 m/s [16]). Propagation delays of signal onsets, relative to the BM's base, are negligible above CF > 2 kHz and grow above ~1 ms for CF < 500 Hz [16] [17] [18]. The onset latency between BM movement and cochlear microphony (electric potential picked up with a cochlear-implant electrode) is ~3 μs [16].

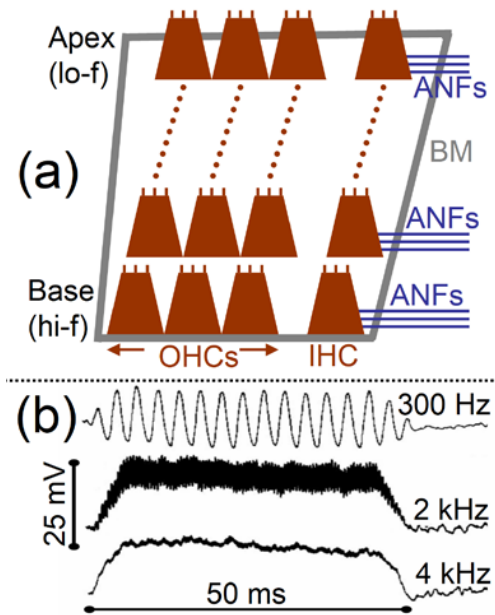

*Fig. 4 (a) Schematic of unfolded basilar membrane (BM) showing tonotopic (tuning by position) arrangement of rows of outer and inner hair cells (OHCs and IHCs). High frequencies (hi-f) resonate closer to the BM's base (near cochlear entrance) and low frequencies (lo-f) at its apex (far end). (b) Analog receptor voltages versus time measured in IHCs of guinea pigs (for 50 ms pure tones with 5 ms ramps) track the stimulus waveform below ~4 kHz but become positive plateaus (independent of stimulus phase) at higher frequencies (data adapted from [19]).*

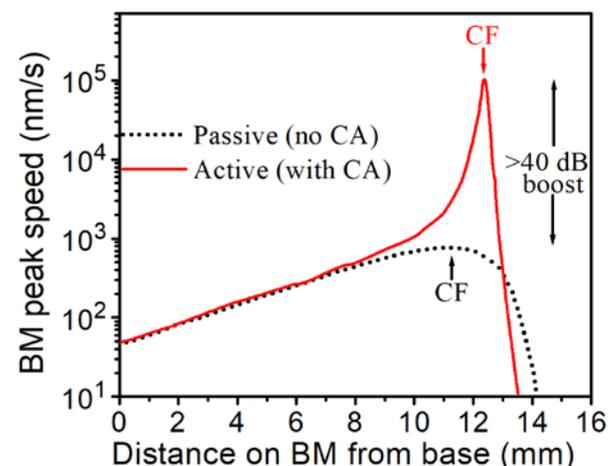

*Fig. 5. Effect of cochlear amplification (CA) on basilar membrane (BM) tuning curves [20]. The tuning curve becomes sharper (improving frequency discrimination) and the characteristic frequency (CF; arrows at which a given BM location oscillates with maximum amplitude) shifts higher. Also there is a >40 dB boost in amplitude and a corresponding reduction in the detection threshold (i.e., increased sensitivity for soft sounds).*

---

[11] Also referred to as a *neural activation pattern* or NAP.

The BM's mechanical tonotopy is augmented by additional gradients along its length in properties such as IHC and stereocilia dimensions, $K^+$ and $Ca^{++}$ influx/efflux times, and ANF conduction speeds and lengths. This gradients-based passive tonotopy is sharpened by an active reinforcement mechanism called *cochlear amplification* (CA) [21] [22] [23] [24] [25] [26] [27]. As shown in Fig. 5, CA enhances the sensitivity, dynamic range, and frequency tuning, and shifts the CF dynamically with level. CA also compresses large BM displacements and protects the ear from loud sounds [25]. Additional frequency sharpening takes place at higher centers due to inhibitory suppression of side flanks of tuning curves.

Although the exact mechanics of CA are still being investigated, it can be approximated by these 3 stages: (1) fast calcium-current-driven OHC hair-bundle motility affecting local TM properties and resonances[12]; (2) voltage-driven[13] OHC somatic motility (involving prestin motors in OHC walls) affecting local BM stiffness and motion; and (3) overall regulation and modulation by neural feedback from higher centers (expounded in a later section). The time frames for these processes are ~15 μs, ~240 μs, and >1 ms respectively [27] [28] [29] [30]. What this means is that CA may not have enough reaction time to operate for brief transients. In this case the analysis of transient signals should not be based on continuous-tone thresholds and parameters, as their conditions are different. At the early onset of a sound, tunings of cochlear filters are broader and have shorter impulse response times, to better evaluate transients [31]. A detailed and mathematical description of cochlear biophysics is given in [29].

### 2.3 Frequency range and hearing loss

Fig. 6 (a) shows how subjective loudness varies with frequency and sound level. The overall shape reflects in large part the acoustical/mechanical spectral transfer function from the external ear up to the BM. Superimposed on this is the tonotopic organization of the ~3500 overlapping CF (IHC) channels. As expected from the gradual left tail of the channel-tuning-curve of Fig. 5, the ELC curves of Fig. 6 do not have a sharp cutoff at low frequencies. On the other hand, at the high-frequency end, there is an abrupt upward divergence in the threshold and in the SPL needed to produce a given loudness sensation. This is related to the sharp cut off on the right side of the channel-tuning-curve of Fig. 5 and the BM tonotopy reaching its highest CF at the basal end of ~16 ± 2 kHz [32]. Thus the functional frequency range for young otologically healthy people is $f_{min}$ = 16 Hz to $f_{max}$ = 18 kHz [33] (or 20 Hz to 20 kHz in memorable round numbers[14]).

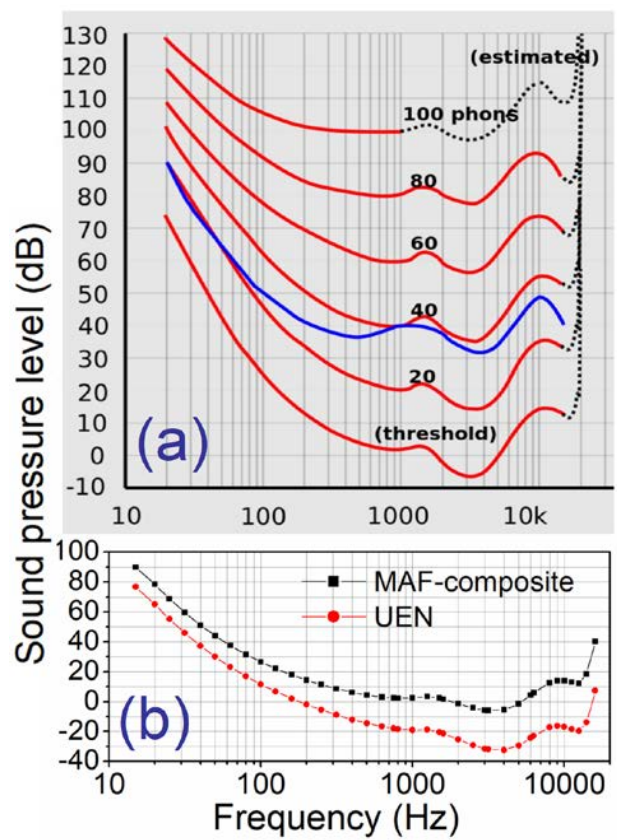

*Fig. 6 (a) Equal (perceived) loudness contours (ELC; red curves) as per the ISO (International Organization for Standardization) 226:2003 standard (revised second edition); the lower 40-phon curve (blue) is for the old ISO standard (first edition) [34]. The lowest contour represents the threshold of hearing. The thresholds for discomfort and pain (not shown) are ~110 and ~120–130 dB at 1 kHz respectively. 1 kHz is taken as the standard frequency at which the loudness in phons equals the physical sound pressure level in dB. The practical human frequency range is 16 Hz–18 kHz, commonly rounded to 20 Hz–20 kHz. (b) A computed threshold curve corresponding to the intriguing concept, developed in [35] [36] [37], of simultaneously stimulating all IHC channels with 'uniformly exciting noise' (UEN). Shown for comparison is the MAF-composite curve, combining ISO226:2003 with ISO389-7:2019 for high-frequency range extension.*

Because of subsequent cross-lateral and cross-frequency neural processing, the binaural threshold of hearing or MAF[15] (minimum audible field) of Fig. 6 (a) is ~3 dB more sensitive than for monaural listening [38] [39]

---

[12] OHC (unlike IHC) stereocilia are attached to the TM.
[13] At ~10 MV/m, the transmembrane electric field is thrice air's breakdown field that gives rise to lightning.
[14] No published result could be found with $f_{max} \geq$ 19 kHz.
[15] In this MAF measurement, loudspeakers placed directly in front of the listener produce a plane wave of one pure tone at a time. The SPL is measured at the position of the head's center with the listener removed. In contrast, MAP (*minimum audible pressure*) measurements employ headphones and the SPL is measured just inside the ear canal. In both cases, listening is binaural-diotic.

[40]. Also it is possible to hear a complex tone whose individual harmonics are below their respective pure-tone thresholds; and if all IHC channels are optimally excited, the calculated effective threshold dips below -30 dB as shown in Fig. 6 (b) [35] [36] [37].

Under exceptional conditions and high sound levels, some individual human subjects have detected frequencies as low as 12 Hz [41] and as high as 28 kHz [42]. More generally, in the animal kingdom, hearing range stretches from 0.5 Hz for pigeons to 300 kHz for the moth species *galleria mellonella*, with some bat species hearing up to 200 kHz [43] [44].

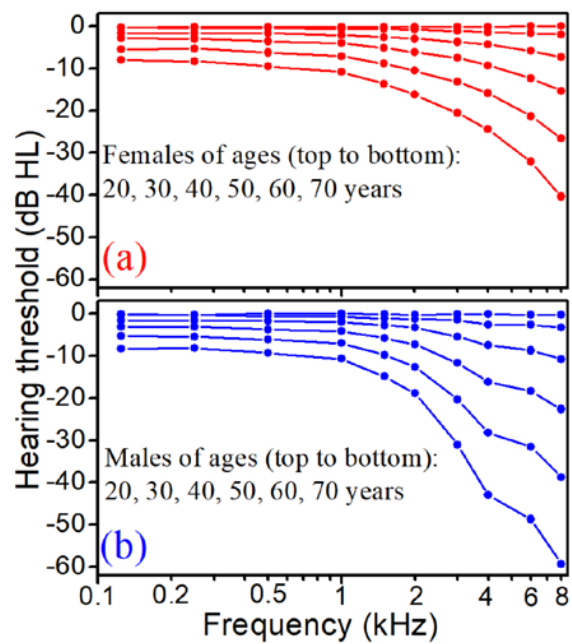

*Fig. 8 Standard ISO 7029:2017 median audiograms showing age-related hearing loss for otologically normal females (a) and males (b) [47] [48].*

Fig. 8 illustrates *presbycusis* or age-related hearing loss. Unlike the notch at speech frequencies caused by noise-induced loss, here there is a progressive bilateral loss of sensitivity starting from high frequencies. On average, females have better hearing than males in humans and are better protected from damage due to loud sounds. This difference is not entirely due to higher environmental noise in historically male dominated jobs, but due to clear biological differences as expounded below.

The hearing losses described above are of the *sensorineural* type, resulting from damage to the hair cells and/or associated nerves, and may be accompanied by *tinnitus* or "ringing in the ears". But hearing can also be compromised by *conductive losses*, in which sound energy is impeded from reaching the cochlea by problems in the external ear (e.g., wax buildup or ear-drum rupture) and middle ear (e.g., fluid accumulation or arthritis of the ossicle joints).

## 2.4 Discrimination of pitch, level, and rhythm

The *just noticeable difference* JND (also called a *difference limen* or DL) defines the threshold change in a parameter that a human can barely discern. JNDs are valuable for evaluating the potential sonic effects of certain distortions. Fig. 9 provides an at-a-glance summary [49] of the classic JND measurements of [50] [51] [52] that are often used for reference. More detailed measurements from another source [53] are tabulated in Table 1. JNDs vary with measurement method and with individuals. [54] and [55] compare the different measurement methods.

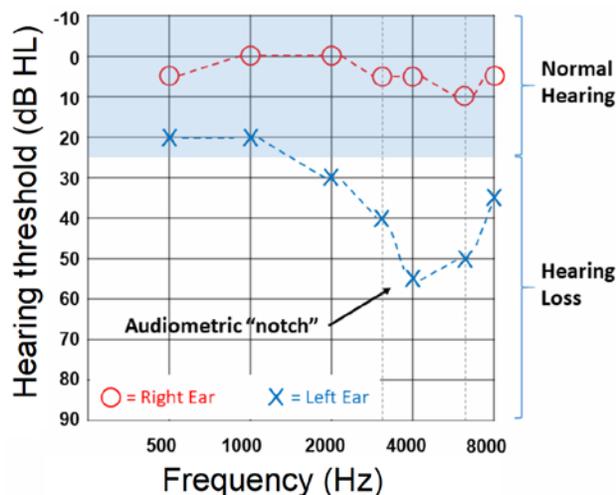

*Fig. 7 Audiogram showing the 'hearing threshold'. This is the difference between the measured minimum audibility level for a particular ear minus a standard 'minimum audibility curve' [45]. Here the right ear is within normal range, but the left one shows a notch characteristic of noise-induced hearing loss. Such 'conventional' audiograms test up to 8 kHz, whereas 'high-frequency' audiograms used in research and for diagnosing age-related hearing loss test up to 20 kHz [46].*

The huge boost in transfer function around the 3–4 kHz speech region (deep dip in threshold in Fig. 6) makes it especially vulnerable to damage by noise exposure[16]. This is evident in the noise-induced notch of a patient's *audiogram*[17] in Fig. 7. The damage occurs primarily to the OHCs (the IHCs are relatively robust) resulting in a reduction in the cochlear amplifier's reinforcing frequency-selective feedback (see Fig. 5 and associated text) as well as its suppressive action against loud sounds. Noise induced loss is thus accompanied by reduced dynamic range, frequency selectivity, and speech discrimination.

---

[16] Cumulative effect of non-specific environmental noise.

[17] Audiograms are usually made under monaural headphone listening conditions.

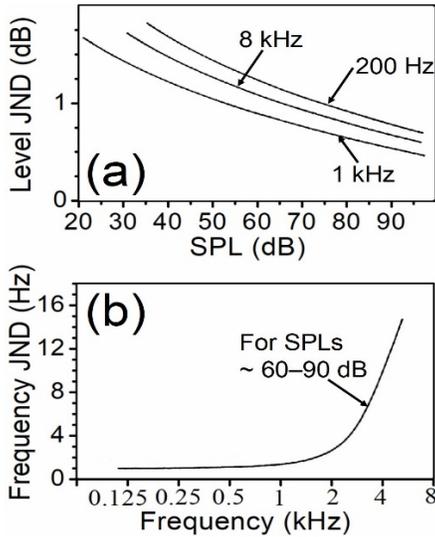

*Fig. 9 Condensed representative curves of just noticeable differences (JNDs) for changes in sound level (a) and frequency (b) for typical musically important levels and frequencies [49] [50] [51] [52]. SPL=sound pressure level.*

Level JNDs are on the order of 1 dB or less over most of the parameter space (SPL > 40 dB and f > 100 Hz), dropping to 0.25–0.4 dB for SPL > 60 dB and f = 1000–4000 Hz. Some early work [56] [57] found, for broadband noise, JND ~ 0.5–1 dB for SPL = 20–100 dB. Estimations have shown that, if information from all 30000 ANFs was used optimally, JND < 0.1 dB should be expected for tone bursts around f ~1 kHz [58].

Frequency discrimination is keenest around 2000 Hz for SPL > 30 dB, where JND = 3 cents[18] or ~0.2% of the pure-tone (single) frequency (some trained musicians can discriminate differences under 2 cents). Notice that this corresponds to a single row of hair cells[19] or less! JNDs tend to be finer when listening with both ears and for complex tones—dropping as low as ~0.1 Hz or ~1 cent [49]—indicating that cross-channel pathways in subsequent neural processing sharpen discrimination beyond cochlear tonotopy. Individuals who are unable to discriminate pitch better than a semitone are said to suffer from *tone deafness* or *amusia*.

| JNDs for LEVEL (in dB) | | | | | | | | | | | |
|---|---|---|---|---|---|---|---|---|---|---|---|
| frequency (Hz) | Sound level (dB SPL) | | | | | | | | | | |
| | 5 | 10 | 20 | 30 | 40 | 50 | 60 | 70 | 80 | 90 | 100 |
| **35** | 9.3 | 7.8 | 4.3 | 1.8 | 1.8 | | | | | | |
| **70** | 5.7 | 4.2 | 2.4 | 1.5 | 1 | 0.75 | 0.61 | 0.57 | 1 | 1 | |
| **200** | 4.7 | 3.4 | 1.2 | 1.2 | 0.86 | 0.68 | 0.53 | 0.45 | 0.41 | 0.41 | |
| **1000** | 3 | 2.3 | 1.5 | 1 | 0.72 | 0.53 | 0.41 | 0.33 | 0.29 | 0.29 | 0.25 |
| **4000** | 2.5 | 1.7 | 0.97 | 0.68 | 0.49 | 0.41 | 0.29 | 0.25 | 0.25 | 0.21 | |
| **8000** | 4 | 2.8 | 1.5 | 0.9 | 0.68 | 0.61 | 0.53 | 0.49 | 0.45 | 0.41 | |
| **10,000** | 4.7 | 3.3 | 1.7 | 1.1 | 0.86 | 0.75 | 0.68 | 0.61 | 0.57 | | |

| JNDs for FREQUENCY (in cents) | | | | | | | | | | | |
|---|---|---|---|---|---|---|---|---|---|---|---|
| frequency (Hz) | Sound level (dB SPL) | | | | | | | | | | |
| | 5 | 10 | 15 | 20 | 30 | 40 | 50 | 60 | 70 | 80 | 90 |
| **31** | 220 | 150 | 120 | 97 | 76 | 70 | | | | | |
| **62** | 120 | 120 | 94 | 85 | 80 | 74 | 61 | 60 | | | |
| **125** | 100 | 73 | 57 | 52 | 46 | 43 | 48 | 47 | | | |
| **250** | 61 | 37 | 27 | 22 | 19 | 18 | 17 | 17 | 17 | 17 | |
| **500** | 28 | 19 | 14 | 12 | 10 | 9 | 7 | 6 | 7 | | |
| **1000** | 16 | 11 | 8 | 7 | 6 | 6 | 6 | 6 | 5 | 5 | 4 |
| **2000** | 14 | 6 | 5 | 4 | 3 | 3 | 3 | 3 | 3 | 3 | |
| **4000** | 10 | 8 | 7 | 5 | 5 | 4 | 4 | 4 | 4 | | |
| **8000** | 11 | 9 | 8 | 7 | 6 | 5 | 4 | 4 | | | |
| **11,700** | 12 | 10 | 7 | 6 | 6 | 6 | 5 | | | | |

*Table 1. Just noticeable differences (JNDs) for various sound levels (listed in boldface along a row of each header) at various frequencies (listed in boldface in the first column) [59],[60]. The top table lists the 'level JNDs' in dB and the bottom table lists the 'frequency JNDs' in cents (1 cent corresponds to a fractional frequency change of $\Delta f/f = 2^{1/1200}$ or 0.058 %). The absence of data at higher SPLs for low and high frequencies stems mainly from the experimental difficulty of producing distortion free signals in these ranges and does not reflect the limitations of the ear.*

---

[18] A cent corresponds to a fractional frequency change of $\Delta f/f = 2^{1/1200}$ or 0.058 %, which is a hundreth of a semitone and twelve-hundreth of an octave. The Weber-Fechner law—that the fractional just-noticeable stimulus change is constant—holds only approximately and holds only near the middle of each range.

[19] Per octave, there are ~400 IHCs and 12 musical semitones (i.e., 1200 cents). Thus a JND of 3 cents corresponds to 3 x 400/1200 = 1 IHC per JND.

Over the ~10 octaves[20] of hearing, pitch can be distinguished only for the middle ~9 octaves; the extreme frequencies fold into the inner CF channels. Hence the musical range (substantially represented by the standard 88-key piano—from $A_0$ = 27.5 Hz to $C_8$ = 4186 Hz[21]) is a subset of the audible range. In total, humans can differentiate ~5000 shades of pitch over the entire audible range and ~1000 over the musical range. Varying both the level and frequency, approximately 330,000 distinct pure tones can be distinguished monaurally [61] [62].

The JND for rhythm is the longer of 2.5 % or 6 ms for note duration and placement, and the JND for tempo is 4.5-8.8 % [63] [64] [65]. Typically, even a rudimentary audio system can well reproduce pitch, level, and duration. Hence their underlying neurophysiology will not be elaborated upon here. The interested reader can explore the aforementioned references.

As discussed earlier, one should be cautious about applying continuous-tone JNDs for analyzing transient stimuli that are too brief to invoke CA action.

### 2.5 Critical bands, ERBs, and masking

The finite spread of the tuning curve, as shown in Fig. 5, has two consequences. A pure tone will excite a *critical band* (CB) of overlapping CF channels whose tuning curves have at least some response to that frequency. Summing over these channels can provide a better *signal-to-noise ratio* (SNR) for measuring the level for a single frequency (the specific neural circuitry that conducts this moving average is detailed below).

Secondly if one frequency falls within the CB of another, the former can have a masking effect on the latter [66]. As a result, in audio, extraneous signals resulting from distortions or noise can be objectionable not only due to their own annoyance value, but because of their tendency to mask low level details that are part of the music. One such low-level sound that is critical to depth perception (see below) is the original reverberation. Indeed, audiophiles claim a greater perceived soundstage depth when noise is reduced, for example through power conditioners or better shielded cables[22].

Modeling the peripheral auditory system as a bank of band-pass *auditory filters* and the CB concept dates back a century [67] [68]. The critical bandwidth is defined between the two points on the skirts (see Fig. 5) where energy, power, and intensity[23] are down by 3 dB or a factor of 2 (speed and displacement amplitudes are down by √2). A convenient quantitative alternative for describing the CB is the *equivalent rectangular bandwidth* (ERB), which is the width of a rectangular bandpass filter with the same power transmission as the actual tuning curve. CBs and ERBs respectively range 10–15 % and 11–17% of the CFs [69]. Each ERB corresponds to roughly a quarter of an octave in pitch. It occupies a distance of 0.9 mm on the BM and includes ~90 rows of hair cells. The ERB (in Hz) for young people with normal hearing can be approximated by [8]:

$$\text{ERB} = 24.7 \, (0.00437 \, \text{CF} + 1) \qquad (2)$$

Further information on this topic can be found in [70].

### 2.6 Heterodyne detection of ultrasound

An individual's $f_{max}$ and the MAF curve of Fig. 6 represent the threshold for *pure tones*. Ultrasonic harmonics in complex tones may heterodyne (mix due to the ear's non-linearity) to produce audible intermediate frequencies, which may influence the NEP and become part of the natural auditory experience [71]. To explore this possibility quantitatively, we will briefly review the experiments and analyses of [71] and [72] whose experimental arrangements are shown in Fig. 10.

A 7 kHz square-wave tone at a listener level of 70 db SPL was played with and without the first-order RC low-pass filter switched in (Fig. 10(a) for the experiment of [71]) or with the loudspeakers spatially misaligned or not (Fig. 10(b) for the experiment of [72]). The listeners' task was to distinguish between the configurations. The audibility lower bound for the first experiment [71] was τ < RC = 4.7 μs (i.e., $f_c$ > 34 kHz) and for the second experiment [72] was τ < d/v = 6.7 μs, with respective statistics $\chi^2$ = 25.9 (p = 3.6 x $10^{-7}$) and $\chi^2$ = 20.5 (p = 6 x $10^{-6}$) well exceeding psychophysical standards[24].

---

[20] The normal frequency range represents a ratio of 18000/16 = 1125 ≈ $2^{10}$.

[21] In musical note-octave notation, the letter corresponds to the note and the number to the octave. Thus "middle C" is $C_4$ (also written as C4, C(4), or C[4]) i.e., the note C in the 4th octave. The frequency standard is defined by $A_4$ =440 Hz. In the scientific-pitch scheme, $C_{-4}$ = 1 Hz, $C_0$ = 16 Hz (threshold of audibility), and $C_4$ = 256 Hz. The Bösendorfer Imperial Grand piano extends down to $C_0$.

[22] This and other anecdotal claims by audiophiles are often dismissed out of hand, but may be worth investigating through formal research and IRB approved blind listening tests for possible verification and furthering insight.

[23] Refresher: *Energy* in joules (J) measures the capacity to do work; *power* in watts (W) is the time rate of work or transferring energy; and *intensity* in watts per square meter (W/m²) is the concentration of power per area.

[24] In psychophysics, a successful *chi-squared* test (for 1 degree of freedom) requires the *chi-squared value* $\chi^2$ = (C − T/2)²/(T/2) + (I − T/2)²/(T/2) to exceed the *critical value* of 3.86 for which the probability (*p value*) of obtaining the result by random chance is <5%; here T is the total number of trials, C is the number of correct judgments, and I is the number of incorrect judgments. [71] also calculated a *discriminability index* of d' = 2.26, which again well exceeds its *criterion* of c = 0.92.

Let us analyze the result of [71] in some detail ([72] is similar). For audibility purposes, a 7 kHz square waveform consists principally of 7 kHz and 21 kHz harmonics, but only 7 kHz is directly audible since by measurement $f_{max} <$ 18 kHz for all the listeners. However, an audible 14 kHz harmonic can be generated due to the ear's compressive non-linear response [73]:

$$y \propto x - bx^2 \quad (3)$$

where x is the "input" amplitude of the incident sound, y is the "output" amplitude at the cochlear BM, and the constant b ~0.01. Potentially, intermodulation distortion (IMD) in the audio chain [74] can also produce 14 kHz. However, this contamination was ascertained to be negligible by directly measuring the listener-position acoustic waveform and spectrum (Fig. 10(c)).

The low-pass filter alters the phase of the 21 kHz. Then, through interference between the non-linearly produced quadratic tone (14 = 2 x 7 kHz) and difference tone (14 = 21 - 7 kHz), the net 14 kHz level changes by $\Delta L_{14kHz} =$ 1.45 dB (for details see footnote[25]). This is comparable to the relevant JND ~ 1–2 dB (see Table 1 and Fig. 9) and hence should be audible.

On the other hand, the low-pass filtering of the experiment also attenuates existing frequencies' levels by:

$$\Delta L = -10 \log[1 + (2\pi f\tau)^2] \quad (11)$$

giving $|\Delta L_{7kHz}| = 0.18$ dB for the 7 kHz fundamental at the threshold $\tau$ = 4.7 μs. This is much lower than the corresponding JND ~ 0.5–1 dB (Table 1 and Fig. 9), making the heterodyne mechanism a more plausible explanation for the audibility of the $\tau$ = 4.7 μs low-pass filtering[26]. The present experiments used a mild first-order filter that introduced small phase and level changes in the ultrasound; a steeper filter that eliminates the ultrasound altogether would completely remove the 14 kHz, causing a drastic 27 dB drop in an audible component.

The present experiments used a 7 kHz square wave, which has mainly one weak ultrasonic component at 21 kHz. The effect should be more noticeable and have a lower discernable $\tau$ for musical-instrument sounds with copious ultrasound [75] [76]. Thus overall, heterodyne detection provides a plausible need for ultrasonic bandwidth in high-fidelity reproduction of music.

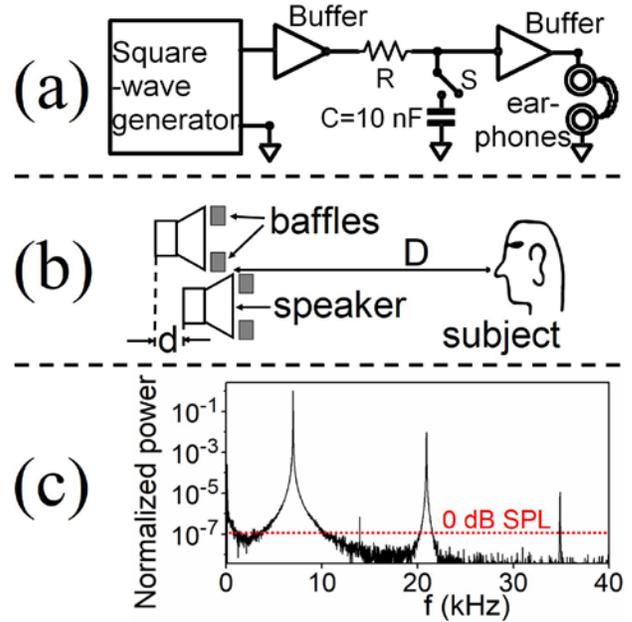

Fig. 10(a) Psychoacoustic experiment [71] proved that a first-order low-pass cutoff frequency $f_c$ = 34 kHz (i.e., time constant $\tau$ = RC = 4.7 μs) is audible with a diotic supra-aural earphones presentation. (b) Psychoacoustic experiment [72] proved that a spatial misalignment d=2.3 mm (i.e., $\tau$ = d/v = 6.7 μs) of loudspeakers is audible to a listener a distance D=4.3 m away. (c) Acoustic power spectrum for [71]; the low relative power of 7 x 10$^{-7}$ (~9dB SPL) of 14 kHz, attests to low intermodulation distortion in the audio chain. Further details can be found in [71] and [72]. C=capacitance; R=resistance; S=switch; v=speed of sound.

In addition to the above mechanism which occurs even at a moderate level of ultrasound (21 kHz at 55 dB), there have been studies demonstrating audibility of high-level (> 85 dB SPL) ultrasound by itself, possibly through the generation of audible subharmonics or due to bone conduction [77] [78] [79] [80] [81] [82] [83] [84] [85].

---

[25] The acoustically measured unfiltered/filtered relative pressure waveforms respectively represented by:

$$P_u = P_0[\cos(2\pi 7000t) + 0.22 \cos(2\pi 21000t+\phi_u)] \quad (4)$$
$$P_f = P_0[0.98 \cos(2\pi 7000t) + 0.18 \cos(2\pi 21000t+\phi_f)] \quad (5)$$

are transformed enroute to the cochlea, by the external and middle-ear transfer function [8], and become:

$$P'_u = P'_0[\cos(2\pi 7000t) + 0.19 \cos(2\pi 21000t+\phi'_u)] \quad (6)$$
$$P'_f = P'_0[0.98 \cos(2\pi 7000t) + 0.15 \cos(2\pi 21000t+\phi'_f)] \quad (7)$$

Non-linear mixing (Eq. 3) converts an input of the form $x = \cos(2\pi f_0 t) + a \cos(2\pi 3f_0 t + \theta)$ into:

$$y \approx \cos(2\pi f_0 t) - b/2 \cos(2\pi 2f_0 t) - ab \cos(2\pi 2f_0 t +\theta) \quad (8)$$

keeping oscillating terms up to $2f_0$ in frequency. The second term (quadratic tone) is phase locked with the fundamental and interferes with the last term (difference tone between $f_0$ and $3f_0$) which is phase locked with $3f_0$, giving a net $2f_0$ (here 14 kHz) amplitude:

$$y_{2f0} = b [\{0.5 + a \cos(\theta)\}^2 + \{a \sin(\theta)\}^2]^{1/2} \quad (9)$$

For our values of b=0.01 (Eq. 3) and a=0.19 (Eq. 6), this amplitude can vary with θ from $y_{2f0} \approx 0.003$ to 0.007 times $P'_0$, i.e., a level range of $L_{14kHz} \approx 19.5$ to 27 dB.

The filtering in the experiment [71] shifts phase by:

$$\Delta\phi = \tan^{-1}(-2\pi f\tau) \quad (10)$$

causing $\Delta\phi_{7kHz} = -11.7°$ and $\Delta\phi_{21kHz} = -31.8°$, i.e. the shift in $\theta = \phi_f - \phi_u = \phi'_f - \phi'_u = 20.1°$. Then Eq. 9 produces up to $\Delta L_{14kHz} = 1.45$ dB depending on the initial $\phi'_u$.

[26] On the other hand it is possible that the standard JNDs are overestimates. In this case the experiments of [71] and [72] provide a more sensitive way to measure them.

## 2.7 Dynamic range and resolution of detail

When comparing the *dynamic range* (DR) between the ear and audio it is important to remember that the information output of the ear is spectrally deconstructed from the outset: first as an array of ~3500 IHC analog receptor potentials and subsequently as an NEP representing the firing rates of ~30,000 ANFs. By contrast a PCM (*pulse-code modulation*) digital sample (or tape magnetization or record-groove modulation in the case of analog) represents the *total* signal for all frequencies combined. The ear's DR is ~100 dB (see Fig. 6) *per frequency* when one pure tone is played at a time, and even higher for broad-spectrum sound. An audio chain—from microphone to playback-system speakers, plus listening room's acoustics and extraneous noise—will be hard pressed to approach the DR of the ear. Also the ear's sensitivity lies within an order of magnitude of the fundamental thermal noise, with a smallest detectable BM amplitude of ~1 pm (picometer) [86] [87] [88]—i.e., a hundredth the size of an atom!

In addition to DR and sensitivity, the vast information contained in the NEP represents an astronomical *resolution of detail* (RD). At a crowded party, we can focus on a single voice being drowned by hundreds of competing sounds, and still notice our name being called amidst the racket—the so-called "cocktail-party effect" [89] [90] [91]. In music, we are aware of the faint reverberation of past notes superposed on the million times more intense currently playing notes; in fact, their ratio serves as a depth perception cue (see below). All this is possible because of a huge RD, which we now estimate from the known DR and JNDs.

Pure-tone JNDs arise collectively from a group of adjacent hair-cell rows, not just one. As a conservative estimate, we will take an entire ERB (~90 channels) as such a group with its DR ~100 dB subdivided by ~100 levels spaced by JNDs of ~1 dB. The frequency range is thus divided into 40 such groups. The NEP can then be thought of as a 40-digit base-100 number that can have $100^{40} = 10^{80}$ distinct values. Even pessimistically estimating each ERB to have only ~10 distinct levels[27] yields RD $>10^{40}$ as a very conservative lower bound. The footnotes[28] show some alternative calculational approaches for estimating RD, which reinforce the above RD $>10^{40}$ lower bound. Even elderly audiophiles who have lost a couple of octaves (i.e., 8 ERBs) of high-frequency hearing (i.e., $f_{max}$ = 4.5 kHz instead of 18 kHz) will have an RD $> 10^{32}$ that is beyond astronomical[29]!

It can be asked how much of this information the brain can actually utilize, and how many times a subtle sonic feature needs to be repeated to form a lasting impression in long-term memory. But even if a fraction of this RD is utilized, it may represent a granularity finer than existing audio systems or measurement instrumentation.

## 2.8 Sound produced by the ear. Masculinity-femininity dependence of OAEs and AEPs.

As discussed earlier, the CA system greatly modifies IHC response characteristics—such as sensitivity and tuning—through active motion of OHCs. This also causes the ear to emit sounds itself (detectable by a microphone inserted into the ear canal) termed *otoacoustic emissions* or OAE [30] [92] [93]. SOAEs (spontaneous OAEs) emit continuously without the presence of external sound and appear as narrow-band peaks on the OAE spectrum. SOAE strength is reflected by the number of peaks. SOAEs are not universal but occur in ~80% of females and ~50% of males. CEOAEs (click-evoked OAEs) are "echoes" produced by the cochlea in response to brief "click" sounds. They can last up to 40–60 ms and their strength is expressed in dB-SPL over a specified bandwidth. Both OAE types are indicative of a properly functioning CA system and are associated with better hearing[30].

A separate measure of auditory function, somewhat related to the CEOAEs, are AEPs (*auditory evoked potentials*) obtained by recording the sequence of brain-wave peaks (through electrodes attached to the scalp) in response to clicks.[31]

Some measurements [30] (see Fig. 11) have shown that OAEs and AEPs decline with decreasing femininity and increasing masculinity as reflected by gender and orientation. It is believed that prenatal hormonal levels (particularly androgens such as testosterone that promote masculinization) harm the CA system during gestation. Gender differences in OAEs are also seen in newborn infants [94], leaving little doubt about the biological basis for auditory gender disparity. It is also interesting that the right ear (for OAEs) or right brain (for AEPs), on average, is more active than their left counterparts (see Fig. 11).

---

[27] Even from a "hardware" point of view, each IHC synapses with ~8 ANFs with different spontaneous firing rates. Besides ANF labeling, individual ANF firing rates also determine sound level. These numbers are *per IHC channel*. Since an ERB's channels are not completely correlated, ~10 distinct levels per ERB is a conservative lower bound.

[28] Another measurement [56] found JND ~ 0.5–1.5 dB over SPL = 5–80 dB for 200 Hz to 8 kHz pure tones; i.e., ~75 steps over 75 dB in DR for 5.3 octaves (~ 21 ERBs). This gives RD = $75^{21} = 10^{40}$ just for this limited subset of frequency and dynamic range. Also see previous footnote.

[29] There are ~$10^{23}$ stars in the observable universe.

[30] OAEs are individualistic like a "fingerprint", and fairly constant throughout life from birth. Due to perceptual adaptation they are not heard by the individual as tinnitus.

[31] AEPs are used to test hearing in people (infants, etc.) who cannot respond behaviorally.

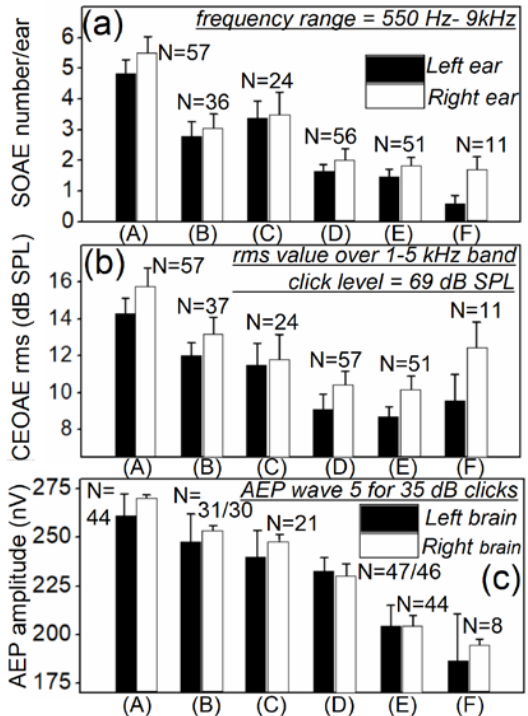

Fig. 11. (a) Spontaneous (SOAE) and (b) click-evoked (CEOAE) otoacoustic emissions, and (c) auditory evoked potential (AEP). N represents the number of participants in each study for each group. Histograms columns (A)–(C) and (D)–(F) respectively represent (hetero-, homo-, and bi-sexual) females and males, roughly following the trend of decreasing femininity and increasing masculinity [30].

## 3 NEURAL PROCESSING IN SUBCORTICAL AUDITORY PATHWAYS

Fig. 12 schematizes the circuitry that processes the ANF signals from the cochlea. The *auditory nerve* AN (a major portion of cranial nerve VIII), contains axons of *spiral ganglion cells* (SGCs of types I and II for IHCs and OHCs respectively) that carry *afferent* (ascending) signals[32]. The AN terminates in the *cochlear nucleus* (CN), where it branches into the *dorsal cochlear nucleus* (DCN), and the anterior (AVCN) and posterior (PVCN) subdivisions of the *ventral cochlear nucleus* (VCN). The trifurcation facilitates parallel processing of three groups of functions as described below [95].

In addition to these afferent pathways, the medial (MOC) and lateral (LOC) *olivocochlear systems* send *efferent* (descending) signals back to the hair cells in the cochlea. The MOC neurons terminate directly on the OHCs, which generate OAEs and mechanical (acoustic) feedback (MF) to the IHCs. This forms one of the functional components of the CA system. The LOC neuron terminals end on the SGC dendrites close to the IHCs and are believed to also sharpen IHC tuning. Further information on the MOC and LOC systems can be found in [27].

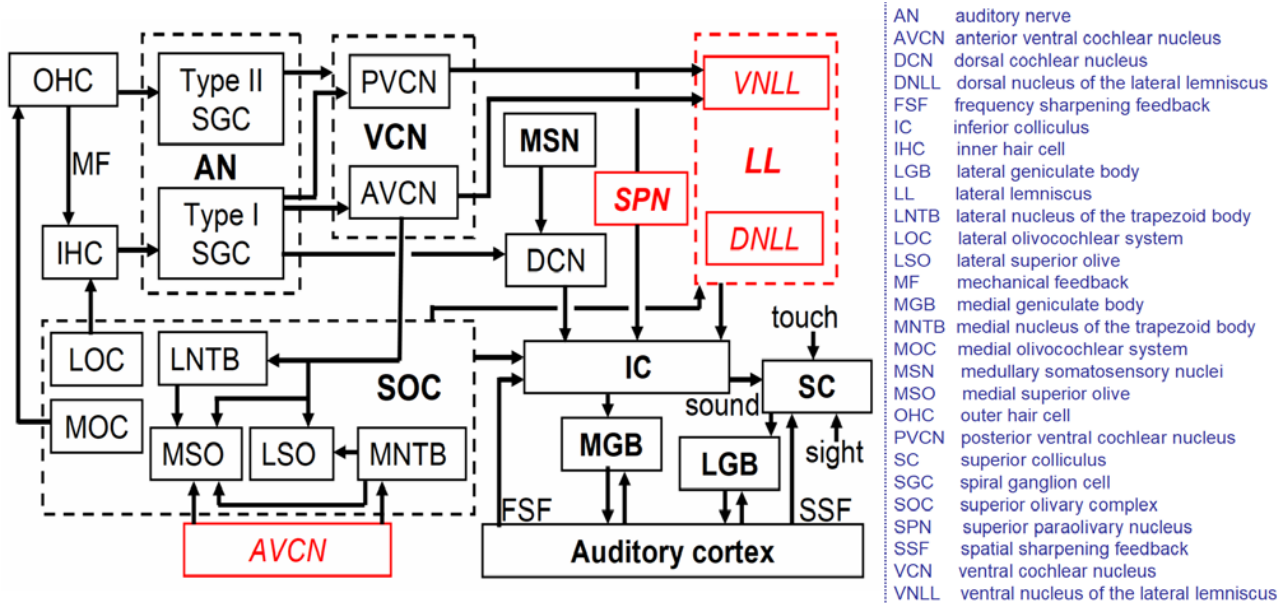

Fig. 12 Simplified flow chart showing some principal auditory neural pathways culminating in the cortex. Contralateral (opposite-side) complexes/nuclei are shown in italicized red. Major complexes and stations are enclosed in dashed-line boxes and labeled in boldface font; subdivisions and nuclei are labeled without boldface. VCN, DCN, SOC, and LL inhabit the 'brainstem' region, IC and SC reside in the 'midbrain', and MGB and LGB are within the thalamus in the 'forebrain'.

---

[32] The firing pattern of the type-I SGCs comprises the NEP "information sample" from which all subsequent conclusions and perceptions are drawn. The type-II SGCs carry correctional feedback from the OHCs that adjusts and fine tunes the CA system.

### 3.1 DCN and elevation localization

One principal role of the dorsal cochlear nucleus is in localization, especially *elevation* (angle in the up-down front-back vertical plane) but to some extent also *azimuth* (angle in the left-right horizontal plane). As mentioned earlier, the spectral transfer function of the external human ear boosts the region of the speech frequencies (roughly as the inverse of the threshold ELC of Fig. 6). Superimposed on this smooth bump are sharp notches and other modulations due to interference of the direct sound entering the ear canal with the reflections from the pinna, head and torso as shown in Fig. 13 (a) and (b). This spectral structure—known variously as HRTF (head related transfer function), ATF (anatomical transfer function), or pinna filtering—varies with direction and can therefore provide localization cues [1] [56] [96] [97]. The measured mammalian HRTF of Fig. 13 (c) shows how the first notch moves up in frequency with increasing elevation for a fixed azimuth [98]. Spectral notch filtering can be employed to artificially manipulate image elevation (e.g., [99]).

The principal neurons in the DCN (particularly the fusiform/pyramidal cells and giant cells) are sensitive to notches and together with DCN interneurons can respond with specificity to complex spectral patterns in stimuli; indeed, cats with lesions in the DCN region are unable to make reflexive responses to sound elevations [100]. The DCN also appears to be involved in other tasks such as suppressing the self-generated sound of our heart beats—the failure of which leads to pulsatile tinnitus [101]—and pathways through the DCN to higher centers are involved in coupling emotional responses to acoustic stimuli [102].

### 3.2 Reflection-delay mechanism for elevation localization

It has been suspected that mechanisms other than HRTF, which are of temporal rather than spectral origin, must also be involved in elevation localization. Humans can localize the elevation of narrowband and low-frequency natural sounds, which cannot be explained by the HRTF spectral mechanism (for f << 3 kHz, the wavelengths are too long for interference).

[103] [104] [105] have proposed a mechanism based on the time delay between arrivals of the direct sound and upper-torso reflections (mainly from shoulders). As illustrated in Fig. 14(a) and (b), the reflection delay increases with elevation: overhead sources entail a round trip from ear to shoulder and back compared to a single trip for forward sources. Being temporal, this model is not specific to a certain frequency range and works down to arbitrary low frequencies. It also works for narrow-band sounds.

Handling of low-frequency information by the shoulder-reflection mechanism appears to be integral in overall elevation localization because it is found that listeners can be confused between front and back directions unless low frequencies below 2 kHz are present [106].

This reflection-delay idea is relatively new. At the present time, it is not known how and where in the brain the delay measurement might occur. However, there are other well studied precedents for delay-measuring neural circuitry (see discussion of the SOC and MSO below). Also indirect corroboration is provided by a fascinating experiment: When an identical sound is played through two loudspeakers positioned along sidewalls directly facing each ear, the sound appears overhead [104] [105] [107] [108] [109]. The explanation given in [104] [105] is that each ear receives two copies of sound: one from its facing speaker and a delayed sound from the opposite-side speaker. The delay for traveling around half the circumference of the head is roughly twice the shoulder-to-ear distance and thus interpreted by the brain as an overhead sound (the apparent elevation drops progressively as the loudspeaker angle is reduced from 180° [sidewalls] to 0° [front wall]).

The above discussion suggests that a recording might capture elevation if made with a wide-polar-response microphone placed distantly (Fig. 14(c)) so as to capture the floor reflection with a delay comparable to a shoulder reflection[33]. This effect was confirmed in [5], where phantom instrument images varied not only in left-right placement and depth, but also in their height. The success of that experiment was aided by well controlled listening-room acoustics, including suppression of floor reflections, to avoid muddling the original recorded reflections.

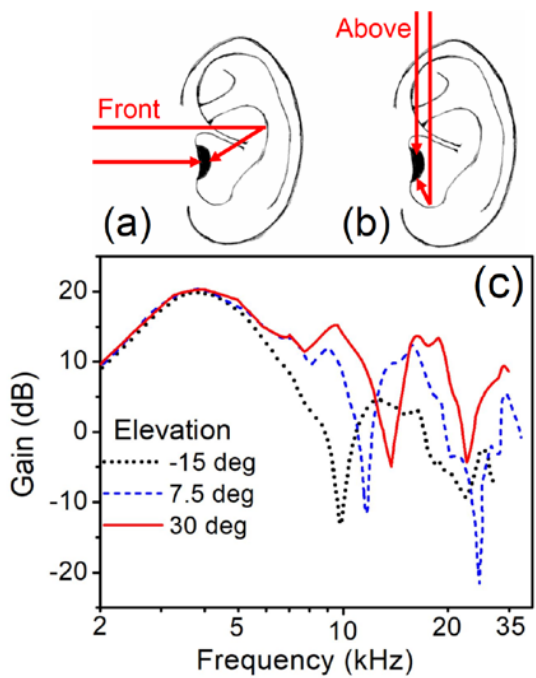

*Fig. 13. (a) and (b): The delays (and hence interference) between direct and pinna-reflected paths depend on the direction of the sound. (c) Measured HRTF (head related transfer function) for different elevation angles of sound direction, comparing the sound level at the eardrum of a cat with the free-field value at the same spatial location in the absence of the cat (based on data from [98]). The important first notch occurs in the 8–17 kHz region. The azimuthal direction of the sound was at 7.5 degrees.*

A related observation is that band limited noise played from a single loudspeaker has an image elevation that increases with the band frequency [110]. Also for a certain range of values, the ground-reflection delay is believed to contribute to depth localization. This is discussed further below.

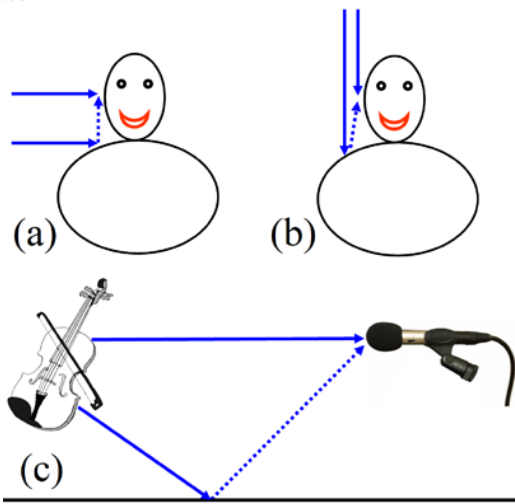

*Fig. 14. (a) and (b) The time delay between direct sound (solid arrows) and reflections from shoulders (dotted arrows) varies with elevation. Unlike HRTF (head related transfer function), the delay-gap cue can work even for narrow bandwidths and low frequencies. (c) A floor reflection captured by a microphone at ~2–5 m distance may, during playback, get psychoacoustically interpreted as a shoulder reflection. Whence instruments will be imaged at different heights.*

### 3.3 AVCN and signal conditioning

*Spherical* and *globular bushy cells* (SBCs and GBCs) in the AVCN refine the timing precision and signal-to-noise ratio of raw ANF signals, through moving-averaging and other processes, before conveying it to higher centers in the brain for further interpretation. SBCs and GBCs (working kind of like synchronous AND gates) respectively combine about 1 to 4 and 4 to 40 closely adjacent ANF signals, preserving the frequency selectivity that started with the BM tonotopy. Inhibitory inputs dynamically reduce the sensitivity at high sound levels, thus requiring a greater number of simultaneous inputs to produce an action potential (spike) [111]. The *endbulbs of Held* between ANFs and SBCs, and *calyces of Held* between GBCs and principal cells in the MNTB (*medial nucleus of the trapezoid body*) represent some of the largest and fastest synaptic terminals in the entire brain. The temporal sharpening of an SBC output compared to its ANF input is portrayed in Fig. 15. (The neurotransmitter kinetics underlying the fast postsynaptic response in bushy cells is discussed in [112] [113].)

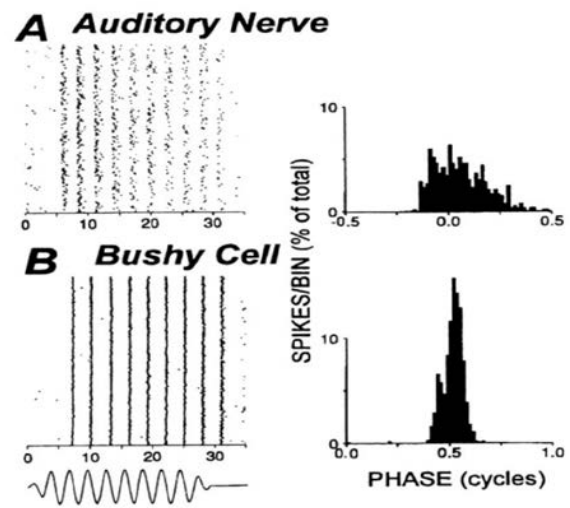

*Fig. 15. Measured spikes corresponding to raw input (row A) and processed output (row B) of a bushy neuron [114]. Left column shows measured action potentials (time is in ms) and the right column shows their phase histograms.*

Besides SBC and GBC, another significant neuron type in the AVCN is the *stellate* (or multipolar) cell. Unlike the bushy neurons that maintain (and in fact enhance) the temporal firing pattern of the ANFs, T-stellates (or type-I stellates or planar cells) exhibit a *sustained chopper* pattern: they fire at a constant rate for the duration of the tone; with the rate having little correlation to the stimulus frequency and phase but instead reflective of the signal strength for its frequency channel. Thus the firing pattern of the ensemble of T-stellates represents the spectrum of the sound. They also encode the envelope— encrypting the onset with high precision [115] [116] as well as rapidly terminating at the sound's offset (the latter arises from inhibitory inputs, which also leads to sideband suppression and sharper frequency selectivity). T-stellates also project to (i.e., feed) the LSO and, along with the SBCs, contribute to the localization process as described below[34].

### 3.4 SOC and azimuthal localization

The SOC (*superior olivary complex*) in the brainstem handles azimuthal localization, through the binaural processes of ITD (*inter-aural time difference*) and ILD (*inter-aural level difference*), which take place in the SOC's two main subdivisions—the MSO (*medial superior olive*) and LSO (*lateral superior olive*) [117]. Going through the VAS (*ventral acoustic stria*), AVCN SBCs project to the MSOs of both sides. SBCs also project to the *ipsilateral* (same side) LSO. An inhibitory input to the LSO arrives from GBCs in the *contralateral* (opposite-side) AVCN, after an inversion in the ipsilateral MNTB. Likewise, the MSO receives ipsilateral and contralateral inhibitory inputs through the LNTB (*lateral nucleus of the trapezoid body*) and MNTB respectively.

---

[33] Instruments at an average height of h~1 m, will mimic 1.5 times the ear-shoulder distance s~0.13 m when a microphone at h~1 m is at a distance d which satisfies the equation: $[d^2/4 - h^2]^{1/2} - d/2 = 0.75$ s, i.e. d≈10 m. This long d also ensures comparable intensities. Typically, microphones are too close to capture height.

[34] Additionally, in the VCN, there are inhibitory *D-stellate* cells (or type II stellate or radial cells), which exhibit an *onset chopper* response that persists briefly after the onset; another VCN neuron is the *small cap cell*. The functions of these neurons is not well understood.

The principal neurons in the MSO are binaural and have a bipolar form, serving as coincidence-detecting synchronous AND gates that fire when signals from the two sides arrive in synchrony [117]. Their nonlinear saturating dendrites make them more likely to fire when both inputs receive signals simultaneously rather than a single large signal at just one input. From some measurements in mammals [118] [119], bipolar cells have an *input resistance* $R_{in} \sim 30$ MΩ, *membrane capacitance* $C_m \sim 70$ pF, and *cell time constant* $\tau_{cell} \sim 2$ ms.

Fig. 16 schematizes the ITD localization process in the MSO [100] [114] [120] (which bears resemblance to the original hypothetical Jeffress model [121]). A bank of MSO bipolar cells are fed signals from the two sides, with graded neuronal delay lines from the contralateral side (Fig. 16(b)). The cells fire increasingly when the acoustic ITD (Fig. 16(a)) is compensated for by a matching neuronal delay. Thus the firing-rate pattern encodes the azimuth. This scheme has been best studied and confirmed in birds; however, there is evidence for the applicability of some of its elements in mammals, possibly augmented by additional mechanisms [122] [123] [124].

Humans can resolve [125] an azimuthal angle of ~1°, which corresponds to an ITD ~ 0.17 sin(1°)/343 ~ 10 μs (per Fig. 16(a)). A more direct approach to measuring threshold ITDs [128] [129] is by playing sound with artificial ITD through earphones[35], for which results are shown in Fig. 17. Below 700 Hz, threshold ITDs decline linearly with frequency, bottoming at 9 μs for 700–900 Hz, and rapidly rising to become immeasurable beyond 1400 Hz at which the wavelength exceeds about 1.5 times the ear spacing d. However, humans can detect the low-frequency envelope of an amplitude modulated high-frequency carrier [126] [127].

Some important observations that emerge from this are: (1) Low frequencies, contrary to myth, can be localized well. In fact, hundreds of hertz are the best frequencies to azimuthally localize (the reason why a time-aligned subwoofer's location vanishes is due to the Franssen effect discussed below). (2) The resolution of time differences has no direct connection with the waveform's period. In fact, the measured ITD=9 μs at f =900 Hz is 123 times shorter than T (=1/f=1.1 ms) and typical neuronal action-potential durations.

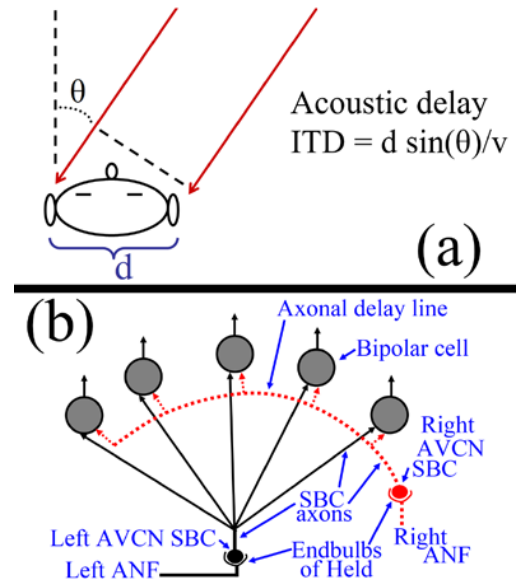

*Fig. 16. (a) Top view of the head showing an off-axis sound (at an azimuthal angle θ) arriving at the far ear with an acoustic delay of ITD = dsin(θ)/v  (here d ~ 0.17 m is the ear spacing and v = 343 m/s is the sound speed). (b) Simplified model for ITD localization in the (left) Medial Superior Olive. Ipsilateral axons (solid black lines) have roughly equal lengths to their target bipolar neurons, whereas contralateral axons (dotted red lines carrying right-ear signals) have graded lengths which compensate for acoustic delays. ANF=auditory nerve fiber; AVCN=anterior ventral cochlear nucleus; ITD= interaural time difference; SBC=spherical bushy cell.*

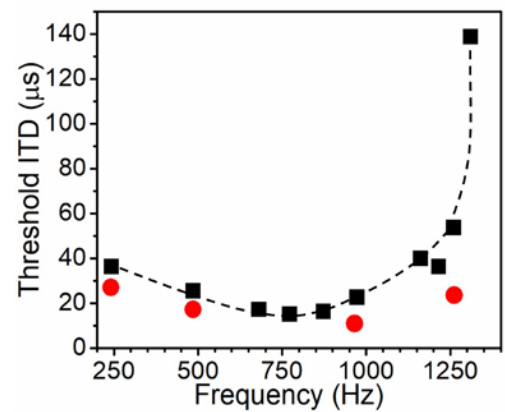

*Fig. 17. Audibility threshold of inter-aural time difference (ITD, in microseconds) versus frequency. Based on data from [128] (red circles) and [129] (black squares with a dashed line as a guide to the eye).*

---

[35] In these experiments, the left and right channels are alternately delayed by Δt. So listeners are actually distinguishing an ITD = 2Δt, which is what is plotted here.

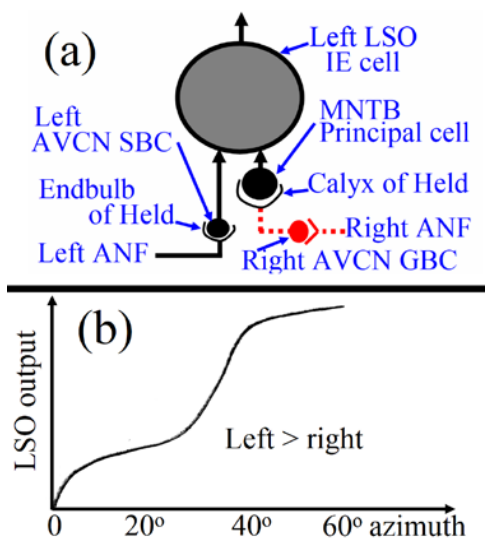

*Fig. 18. (a) Inter-aural level difference (ILD) is measured by the (left) LSO IE neuron by combining the excitatory left signal with inhibitory right signal (inverted in the MNTB cell). (b) The resulting difference appears as the output of the LSO cell (angles are absolute values). ANF=auditory nerve fiber; AVCN=anterior ventral cochlear nucleus; GBC=globular bushy cell; IE= inhibitory-excitatory; LSO=lateral superior olive; MNTB=medial nucleus of the trapezoid body; SBC= spherical bushy cell.*

High-frequency azimuthal localization takes place in the LSO as shown in Fig. 18. The LSO's IE (inhibitory-excitatory) binaural neurons, together with an inversion in the MNTB principal cell for the contralateral signal, act like "NAND gates". Their output reflects the ILD and hence the azimuthal angle. Because long wavelengths diffract around the head, preventing them from casting an acoustic shadow, ILD does not work for low frequencies. Neither ITD nor ILD works well around 1500-2500 Hz where the two mechanisms cross over. Signals from MSO and LSO merge together at higher centers such as the *lateral lemniscus* (LL) or *inferior colliculi* (IC). In addition to ITD and ILD in the SOC, the DCN also encodes azimuth through spectral changes (hence you can differentiate azimuth even with one ear).

### 3.5 Distance (depth) perception

Auditory distance (r') perception is poorer than elevation and azimuthal localization, and has been less researched. Also it is compressed—r' ≈ $r^{0.45}$ where r is the real distance—being overestimated for close sounds and underestimated for distant sounds, and is much less accurate than vision [130].

The first depth cue is the sound level. This falls off at 6 dB per doubling of distance for an omni-directional source in an anechoic room, and slower otherwise. The second cue is the *direct-to-reverberant intensity ratio* (DRR), whose sensitivity maximizes around DRR= 0 dB. DRR appears to work through reverberation's reduction of amplitude modulation (AM), which becomes encoded in the firing rates of IC neurons [131] [132]. The third depth cue is spectral shape, especially for long (>15 m) distances, due to air's greater absorption of high frequencies. Hence loudspeakers that are "bright" (richer highs) tend to image closer to the listener and are referred to as "forward". This spectral mechanism has been confirmed by experiments in which sounds that were progressively low-pass filtered were judged to be more distant in blind trials [133] [134]. Spectral content also provides a cue for judging very short distances (<1 m) due to the diffraction effects of the head [135].

Some other depth mechanisms are binaural cues and HRTF parallax, that pick up changes in ILD, ITD, and average spectrum when the listener turns or moves their head, and also dynamic cues caused by motion of the sound source [130]. Additional suggested depth cues include the initial time delay gap for the first reflection and the shape of the reverberation decay curve [1] [136].

### 3.6 Reflection management and stereo imaging

Other than in an anechoic chamber, direct sound reaching the ear is always accompanied by countless reflections. To avoid overwhelming our awareness, the brain integrates the information so that not every reflection is perceived as a separate event.

Broadly speaking, the brain handles the information in the following way: (1) Early reflections lead to *summing* (or *summative*) *localization*, where direct and reflected sounds are integrated to image at their approximate ''center of gravity" based on the relative delays and intensities. (2) Intermediate reflections lead to the *precedence effect*, in which the location appears predominantly at the leading source. (3) Late reflections lead to echoes (the reflection is perceived as a separate event). The boundaries between the three regimes are not clear cut and depend on details such as the level and type of sound ([1] [56] [137] provide further details). But as a rough guide, one can take "early" as below ~1 ms and the boundary between "intermediate" and "late" as ranging from ~5 ms for impulsive sounds up to ~40 ms for speech or music[36].

The various auditory mechanisms play different roles in stereo imaging versus natural localization. Stereo has only two actual physical sources and azimuthal differentiation occurs through summing localization (recordings typically encode just the inter-channel intensity difference). Whereas when listening to a live ensemble, the azimuths of various instruments are differentiated mainly by ITD and ILD rather than summing localization. Similarly the HRTF mechanism shouldn't work for an (unmanipulated) stereo recording[37]. Thus the virtual soundstage created in stereo cannot be expected to exactly match the original spatial scene no matter how accurate the audio system; although the order of placement (e.g., instrument A is to the left, above, and behind instrument B) might be reproduced, albeit with diminished and distorted dimensions.

---

[36] Even when the reflection is not perceived as a separate event, it still alters the percept of the sound.

An interesting variation of the precedence effect is the *Franssen effect*, whereby the perceptual location of a source latches onto the leading transient as demonstrated in the following experiment [138] [139] [140]: A pure tone with a sharp onset is played from (say) the left speaker while its power $P_{left}$ is exponentially faded out ($P_{left} = P_0\exp\{-t/t_0\}$). The right speaker is then gradually faded in ($P_{right}=P_0[1- \exp\{-t/t_0\}]$) and is kept on for a long time $\Delta t$; the total power ($P_{right}+ P_{left}$) is constant. At the end, the reverse transitions are applied. The listener always perceives the sound to come only from the left speaker for the conditions $\Delta t <$ 4s for $t_0 <$ 40 ms (Franssen effect F1) and $\Delta t \sim \infty$ for $t_0 \sim$ 15 s (effect F2). This underscores the importance of the onset and offset transients (attack and decay). Their crucial roles in pattern recognition and timbre are discussed below.

### 3.7 PVCN and VNLL: Pattern recognition and transient resolution

The "where" aspect of sound—localization—is encoded by the circuitries of the DCN and SOC as discussed above. The "what" aspect—pattern recognition—is based on spectral and temporal fine structure, whose extraction begins in the brainstem nuclei and is then integrated in *ventral nuclei of the lateral lemniscus*[38] (VNLL) before being forwarded on to higher centers such as the IC [141]. Encoding of the spectrum begins in the T-stellate cells in the AVCN and is involved in the identification of vowels and musical-instrument formants. (Monaural) encoding of onset transients (attacks)—which contribute to instrumental timbre and consonant differentiation[39] [142]—begins with *octopus cells* (OCs) in the PVCN (*posterior ventral cochlear nucleus*).

Both binaural ITD and monaural TR (*transient resolution*) involve synchronous AND gating—whereby convergence of signals reduces jitter and leads to extraction of temporal information that is a fraction of the involved periods [114] [117] [119] [141] [147] [150] [151] (also see Fig. 15, Fig. 16, Fig. 17, and their associated discussions). ITD encodes synchronicity between left and right sides per frequency and TR encodes synchronicity between onsets of different frequencies per side (the narrower the impulse or attack, the closer in time the activation of different frequency channels will be). A quantitative estimate for TR can be obtained, based on the established ITD ~10 μs value, by comparing TR and ITD neural circuitries.

We first review the neuronal action-potential process. The electric potential V of a neuron is mainly controlled by the influx/efflux of $Na^+$, $Cl^-$, $Ca^{++}$, and $K^+$ ions through channels (gates) that are activated mechanically, electrically, or chemically. When an ANF fires, it releases the neurotransmitter glutamate into its synapse. This binds with chemically-controlled sodium ($Na^+$) gates on the postsynaptic (target) neuron, causing an inflow of $Na^+$ ions. This depolarizes the neuron (V increases above its resting value of about -70 mV) producing an EPSP (*excitatory postsynaptic potential*). When multiple ANFs synapse onto a single target cell, their EPSPs add (if they concur in time) and generate an action potential if V > -55 mV. Then voltage-controlled sodium gates open, further increasing V to ≈+40 mV. With some delay voltage-controlled potassium ($K^+$ outflow) and chloride ($Cl^-$ inflow) gates open, hyperpolarizing V to ≈ -90 mV. During the ensuing refractory (resetting) period, the neuron tends to ignore input spikes or has a higher threshold.

For a synapse receiving an inhibitory input, a neurotransmitter such as glycine binds to a $Cl^-$ or $K^+$ gate which reduces V (hyperpolarization) causing an IPSP (*inhibitory postsynaptic potential*). A neuron fires if the summation of all the IPSPs and EPSPs occurring within an *integration window* $\Delta t$—related to the *cellular time constant* $\tau_{cell}$, ion influx/efflux/leak times, refractory period, etc.—pushes the net V above the -55 mV threshold.

Relative to the acoustic signal, ANFs will have an initial temporal variability that can be represented by a Gaussian probability-density function:
$$g(t) = t_0^{-1}[2\pi]^{-1/2} \exp(-[t/t_0]^2/2) \qquad (12)$$
The probability that a target neuron with a rectangular window $\Delta t \ll t_0$ will fire upon receiving excitatory spikes from N ANF channels synchronously within $\Delta t$ is:
$$p(t) \approx (\Delta t/t_0)^N [2\pi]^{-N/2} \exp(-[t/\{t_0/\sqrt{N}\}]^2/2) \qquad (13)$$
Notice that the temporal spread got sharpened from $t_0$ to $t_0/\sqrt{N}$ (sharpening will be less if $\Delta t \ll t_0$ doesn't hold). And we see from Fig. 15 for SBCs, that after a convergence of just N ~ 4, over 15 % of their output spikes fall within a single histogram bin of ~3 ms or ~1% of a period. In the binaural ITD process, two such AVCN SBC outputs converge in an MSO bipolar neuron (i.e., 8 total convergences) resulting in a threshold ITD ≈ 10 μs (Fig. 17). A more detailed mathematical description of neuronal spikes, and their firing rates along with input-output correlation functions can be found in [143].

In the monaural TR pathway, OCs (which are exquisitely better adapted for timing determination than SBCs) converge far more ANF signals (N>60 instead of N~4) and there is bank of ~200 OCs [144] [145]. 4 OC outputs converge in neurons[40] of the VNLLv (*ventral subdivision of the VNLL*) compared to just 2 SBC outputs converging in the MSO principal cells for ITD. Thus the total convergences that go into the $t_0/\sqrt{N}$ expression are N~240 instead of ~8. This leads to TR ~ITD/$\sqrt{[240/8]}$ ~2μs (the actual value could be higher because the Δt << $t_0$ condition is not exactly satisfied). But besides the higher N, OCs are superior to the other 3 relevant neuron types by at least a factor of 2 by every measure (see Fig. 19, its caption, and the accompanying footnote), and there will be a further lowering for two-ear dichotic listening [56] relative to this single-ear monaural estimate. Thus based on these physiological comparisons, the ultimate monaural TR can be expected to fall roughly in the ~1–10 μs range. This agrees with the measured ~4–10 μs TR thresholds for discriminating the gap between double pulses [146], which is the only relevant experiment that could be found in the literature (as discussed below, various other "temporal resolution" experiments do not correctly probe "transient resolution" as defined here). Note that TR has no direct connection with $f_{max}$. Thus high-frequency hearing loss will not compromise the synchronicity detection between frequencies that are still audible.

---

[37] Depth in stereo, or even mono, occurs by some of the same mechanisms (intensity, DRR, and spectrum) as in natural localization.

[38] The VNLL has a largely monaural function and is fed by the contralateral VCN. The DNLL is fed by the ipsilateral MSO, LSOs of both sides, and contralateral DNLL.

[39] Patients with otherwise normal audiograms have deficits in speech recognition when they have low ANF synchrony, e.g., due to auditory neuropathy [95].

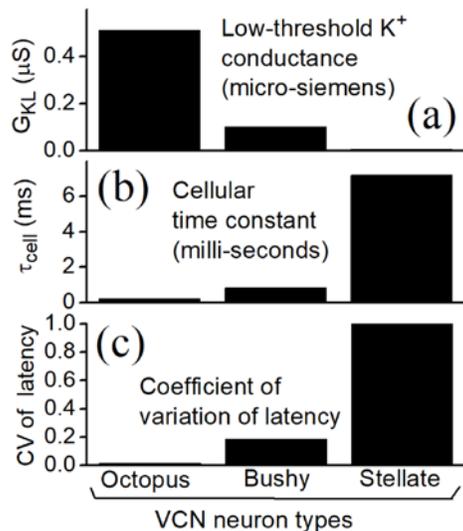

*Fig. 19. Comparisons between octopus, bushy, and stellate neurons of the VCN (ventral cochlear nucleus) [41]. $\tau_{cell}$ ~ $R_{in} C_m$; where $R_{in}$ and $C_m$ are the input resistance and membrane capacitance. MSO (medial superior olive) bipolar neurons have $\tau_{cell}$ ~1-3 ms. $R_{in}$ is ~6 MΩ for octopus neurons, ~70-75 MΩ for bushy and stellate neurons, and ~20-75 MΩ for the MSO neurons. Ordinate symbols and units are explained in each panel. Based on information from [147] [148] [149] [150] [151] [152].*

---

[40] These VNLLv neurons resemble VCN SBCs in their shape and in receiving (OC) inputs through endbulbs.
[41] A strong $G_{KL}$ facilitates constant latency and brevity of synaptic responses; a short $\tau_{cell}$ reduces the time window for integrating synaptic currents from different dendrites

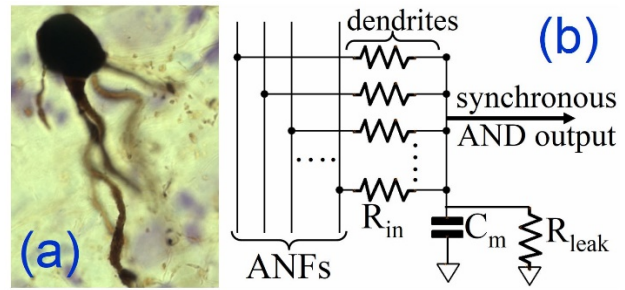

*Fig. 20. Octopus neuron of the PVCN (posterior ventral cochlear nucleus). (a) Optical micrograph. (b) Equivalent circuit. Only 4 of the ~60 ANF (auditory nerve fiber) inputs are shown. $R_{in}$ and $R_{leak}$ are the input and leak resistances, and $C_m$ is the cell-membrane capacitance.*

Fig. 20 shows an image and functional diagram of an OC. Its dendrites are arranged "perpendicular" to ANFs so as to assess the synchronicity of wide ranging frequency channels (whose simultaneous excitation is higher for a narrower impulse); this is in contrast to the SBC's dendrites arranged "parallel" to ANFs so that they perform a moving average of closely adjacent frequency channels. OCs have a leaky cell membrane ($R_{leak}$) to shunt spontaneous currents and tighten Δt (EPSPs leak away quickly, thus putting a higher demand on synchronicity of inputs). OCs produce a single sharply timed response at the onset of tones that are loud enough to excite enough of its ANF inputs to exceed threshold. The exceptionally thick axons of OCs conduct faster than bushy and stellate cells, resulting in shorter latencies [141] [153].

### 3.8 Phase, frequency, and time

Phase and frequency are quantities most meaningful for *periodic* signals or waves. For interference between a loudspeaker's direct sound and floor reflection delayed by Δt, the relative phase Δϕ (in radians) is related to Δt:

$$\Delta\phi = 2\pi f \Delta t = 2\pi \Delta t/T \quad (14)$$

But for *impulsive* sounds that don't overlap or interfere, it is meaningless to apply Eq. 14 and talk about a phase difference. Similarly, frequency bands of time-misaligned drivers in a loudspeaker have a well-defined relative time delay (independent of frequency within each band) but not a constant phase shift. In physics or engineering, the characteristic time of a periodic signal is often taken to be T=1/f or 1/2πf. But we saw earlier that the temporal discrimination by the auditory system can be 2 orders of magnitude better.

Any *signal* can be represented as either a time-domain waveform V(t) ("oscilloscope view") or a frequency-domain spectrum $V^*(f)$ ("spectrum-analyzer view"). Both (frequency channels); neuronal-response latency reflects the delay between the stimulus onset and cell response; and CV (coefficient of variation) of latency reflects its jitter (lower values provide better timing precision) [141][151].

have equivalent information and are transmutable through the Fourier transform/inverse-transform. However, a system's *response* (i.e., transfer function between input and output) is *not generally transmutable* between the time and frequency domains. It is transmutable only for the restricted case of a linear and time-invariant system, which applies neither to audio components nor the ear, since their responses depend on the type and level of the signal and its history. As a result, it is not possible to deduce the exact transient response from the spectral transfer function or other measurements using continuous signals.

During the silence before a sound's onset, the cochlear response is primed for broadband (lower black dotted curve in Fig. 5) transient detection by the PVCN-VNLL pathway. During steady sound, the cochlear response is modified by CA action—trading frequency selectivity for impulse response (upper red curve in Fig. 5). Hence experiments that involve gaps in noise or tones [154] [155], or special temporal structures such as iterated ripple noise [156], assess some form of auditory temporal capability but not its transient resolution that is relevant for timbre as explained below. Measurements such as [154] [155] [156] are irrelevant for audio, as there are no such distortions in practice. Also experiments that discriminate between ordering of short and tall pulses [157] [158] are not evaluating the TR mechanism as described above, which measures the temporal proximity of the onsets of frequency components *regardless of their ordering*.

Musical notes are characterized by four principal attributes: (1) pitch (perceived periodicity), (2) duration, (3) loudness (perceived sound level), and (4) timbre (tonal quality or color). Achieving realistic sound levels, pitch, and duration is less challenging than reproducing convincing timbre. While the spectrum is crucial in determining pitch, it is not as omnipotent in determining timbre. A musical instrument's resonator (sound box) and the air cavities in the vocal apparatus have broad resonant peaks at certain frequencies called *formants*. Formants shape the spectrum, i.e., relative powers of harmonics[42]. Frequency-response irregularities in audio alter these formants and potentially the timbre. However, as seen in the earlier section on JNDs, harmonic powers typically need to change by >0.2 dB (i.e., ~ 5%) to be audible. The FR of most HEA components is more stringent than this. Thus, besides adding noise, differences in sound quality at the level of HEA likely result from time-domain alterations.

Although counter intuitive, a change in a complex tone's waveform shape caused by shifts in relative harmonic phases is largely inaudible since, to first order, the NEP doesn't directly access the waveform itself, but only the spectrally decomposed information of the IHC channels. This assertion of phase deafness is called *Ohm's law of acoustics* [159] [160] and holds well for a repetition rate (implied fundamental frequency) above 400 Hz and when only few and low harmonics are present (i.e., a waveform closer to a pure tone and less spiky). Phase shifts can be detected [161] in a complex tone with a low repetition rate (e.g., ≤ 125 Hz) and numerous in-phase harmonics (e.g., at least the first 12), where the waveform resembles widely separated sharp spikes and is therefore detectable through the TR mechanism.

Thus frequency and phase distortions in HEA are less likely to harm timbre compared to temporal factors such as: (1) waveform envelope (with its principal stages of attack, decay, sustain, and release); (2) different buildup rates/onsets of harmonics; and (3) transient noises such as clicks from picking[43]. This importance of the temporal onset/offset of a note has been strikingly demonstrated in a classic experiment [162] in which various wind instruments were recorded and then played with the beginnings and ends of the notes marginally clipped off, so the spectra hardly changed. The professional musicians had difficulty recognizing their everyday familiar instruments as illustrated in Table 2. [163] [164] [165] discuss temporal and other factors involved in stream segmentation.

| Actual instrument ▼ | Listener judgments | | | | | | | | | |
|---|---|---|---|---|---|---|---|---|---|---|
| | Flute | Oboe | Clarinet | Tenor sax | Alto sax | Trumpet | Cornet | French horn | Baritone | Trombone |
| Flute | 1 | 2 | | 1 | 6 | 5 | 4 | | | 4 |
| Oboe | | 28 | | | | | | | | |
| Clarinet | 1 | 1 | 20 | 4 | 3 | | | | | |
| Tenor saxophone | | | 25 | 2 | 1 | | | | | |
| Alto saxophone | | | | 3 | 4 | | 1 | 11 | 5 | 5 |
| Trumpet | 8 | | | | 6 | 2 | 2 | 4 | 1 | 3 |
| Cornet | | 1 | | | | 12 | 15 | | | |
| French horn | 1 | | | 2 | 3 | | | 5 | 6 | 6 |
| Baritone | | | 1 | 1 | 2 | 3 | 2 | 4 | 7 | 3 |
| Trombone | 2 | 1 | | 5 | 3 | | | 1 | 5 | 9 |

*Table 2. The classic "confusion matrix" experiment by Berger [162]. Clipping off the beginnings and ends of notes makes instruments hard to recognize. Thus onset and offset transients, and small changes in the envelope, greatly affect the timbre. Spectral formants alone aren't adequate for instrument identification.*

In psychoacoustic studies, there is some interest in determining whether the root of audible discernment is "spectral" or "temporal". From a signal point of view, as discussed above, there is no fundamental distinction. In the auditory system, spectral would imply differences in the instantaneous NEP and temporal would be related to the NEP's time evolution. But a change in signal affects both aspects, which are simultaneously parallel processed by separate neural pathways starting at the brainstem.

What matters pragmatically is the maximum allowable *temporal smear* τ in an audio chain that has no audible effect. The quintessential experiment for this is [146], which compared a pair of 10 μs pulses separated by a space

---

[42] In speech, the positions of the jaw, tongue, and lip shift the formants to produce the different vowel sounds.

[43] Vibrato and tremolo (undulations in frequency and amplitude) and steady noises (bowing, hiss of blown air, etc.) are some other factors that influence timbre.

Δt versus a single 20 μs pulse. This produced a discernability of Δt ~ 10 μs when the stimuli were isolated and Δt ~ 4 μs when they were repeated with a periodicity of 0.2 ms. [146] was inconclusive as to the spectral versus temporal basis of the discrimination, and it correctly pointed out (first sentence on their page 464) that JNDs measured with continuous tones cannot be quantitatively applied to analyze transient signals.

[166] [167] probed another temporal alteration that is relevant to (digital) audio, which is jitter. The stimuli were pulse trains with temporal perturbations Δt in the interpulse intervals. They found a discrimnation threshold of Δt ~ 0.1 μs. Here again there was no concrete conclusion regarding the temporal versus spectral basis for the discernment. These various older experiments are worth repeating using modern intrumentation (the TDH-39 headphones used in [146] had $f_C < f_{max}$) and analyzing the results in light of current auditory knowledge.

### 3.9 Time-frequency uncertainty principle

An interesting principle that comes up in discussions of temporal resolution is the Fourier uncertainty relation, which limits the product of the simultaneous precisions Δt and Δf, for time and frequency respectively, to:

$$\Delta t \, \Delta f \geq 1/4\pi \quad (15)$$

where Δt and Δf are the standard deviations of their respective normalized distributions. The minimum uncertainty product holds for a packet with Gaussian envelope where the waveform and its spectrum have the respective probability distributions:

$$P(t) = P_0 \exp(-t^2/2[\Delta t]^2) \cos(2\pi f_0 t) \quad (16)$$
$$P(f) = P_0 \exp(-[f - f_0]^2/2[\Delta f]^2) \quad (17)$$

It is known that Eq. 15 holds for linear operations in time-frequency analysis [168] but not for non-linear operations: e.g., measuring the temporal spacing between zero crossings within the packet can provide exact information about the $f_0$ in Eqs. 16 and 17. A similar non-linear analysis occurs in the auditory pathway through PVCN and VNLL where transient discrimination is based on direct onset timings rather than spectral analysis. It is therefore no surprise that the hearing mechanism can considerably beat the uncertainty principle (i.e., the Eq. 15 limit), as has been demonstrated experimentally [169] [170].

### 3.10 Bandwidth and time-domain behavior in audio

Based on what was discussed above about auditory TR and the factors influencing timbre, it is clear that an audio system's time-domain behavior—especially that which affects the onsets/offsets of sounds—will influence its fidelity, as has been stressed by several authors [171] [172] [173] [174] [175] [176]. There are distortions of various origins that can affect the edges of a signal such as cascaded reflections that add oscillatory tails (e.g., Fig. 8 of [177]), residual decays due to non-ideal capacitive behavior (e.g., Fig. 6 of [177]), uncontrolled impulse response with overshoot and ringing, etc. But fundamentally, every audio component/system is a low-pass filter with a finite *cutoff frequency* $f_c$ (-3 dB upper bandwidth limit) and consequently has a finite *temporal smear*[44] (e.g., [171] [172]):

$$\tau \sim 1/f_c \sim 1/f_s \quad (18)$$

where, in the case of a digital system, $f_s$ is the sampling period. In a detailed analysis, Eq. 18 will be modified by additional time-domain distortions, some of which were mentioned above.

One measure of this smearing is the shortest time gap between two impulses that can be resolved separately rather than merged as a single overlapped impulse. This is akin to how a telescope's (angular) resolution is defined: Fig. 21(a) shows the "Airy pattern" point-spread-function of an ideal telescope, representing an angular spread (Raleigh criterion) of $\theta_C=1.22 \, \lambda/d$ radians (d= aperture diameter and λ= wavelength). As shown in the profiles of Fig. 21(b), two stars closer than $\theta < \theta_C$ get blurred together. Note that this is independent of the pixel density and bit depth of the imager—both can be infinite and the stars would still blur together. Similarly, the finite bandwidth of an audio component limits the sharpness in time with which a peak can be defined.

Fig. 21(c) shows the measured audio output waveform from a DAC[45] fed a 16 bits/44.1 kHz wave file with a single sharpest possible spike (all samples are zero except for one sample of maximum amplitude [$2^{16}$–1]). It looks similar to the profile of the Airy pattern. The spread in time has a measured full-width-half-maximum FWHM = 29.2 μs and a 90% to 10% fall time of 14.7 μs, both comparable to $\tau \sim 1/f_s$ =22.7 μs in agreement with Eq. 18. τ limits the closest separation of two spikes independent of the bit depth[46].

In the literature (e.g., [178]) one finds the following alternative definition of temporal resolution:

$$\tau^* \sim 1/[2^N f_s] \quad (19)$$

This $\tau^*$ represents the smallest time shift of a waveform that can be detected as a different digital value, not how narrowly in time an impulse can be represented. To distinguish it from the *temporal smear* τ, we will refer to $\tau^*$ as the *time-shift discrimination*.

---

[44] The prefactor in Eq. 18 depends on the sharpness of the cutoff; $\tau = 1/2\pi f_c$ for a first-order low-pass filter.
[45] Muse Audio USB Mini DAC (other DACs and CD players tested differed in detail but had comparable FWHMs).
[46] However, it is tied to $f_s$ through the anti-aliasing and reconstruction processes.

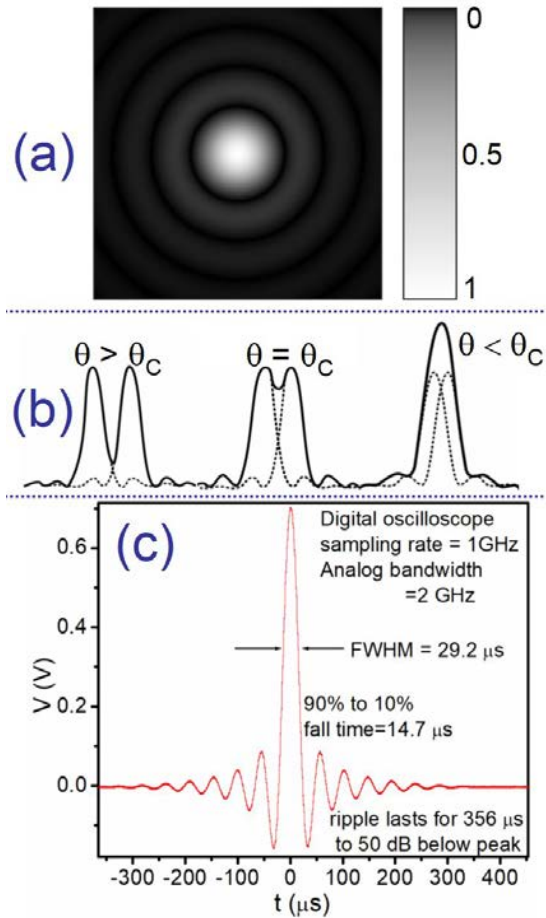

*Fig. 21. (a) The point-spread-function of a telescope of finite aperture (side bar indicates the normalized image illuminance). (b) Two sources (e.g., stars) with angular separation θ < θ_C (Raleigh criterion) merge together, regardless of the imaging pixel density or bit depth. (c) Measured output-voltage waveform from a DAC (digital-to-analog converter) for a unit-sample impulse.*

The temporal response of an audio system concerns more than just resolving transients. There is also the matter of the decay's *cutoff time* $\tau_c$ (typically $\tau_c \gg \tau$) taken for a signal to drop to an undetectable (e.g., system noise) level. Remembering the ear's phenomenal dynamics of DR > $10^{12}$ and RD > $10^{40}$, it is clear that common engineering and physics fractions (such as $1/e = 1/2.72$ for an exponential decay or 90%-to-10% fall) will underestimate how long residue from past sonic events will linger and contaminate subsequent sound. Measuring the extended decay over $t \gg \tau$ and $V \ll V_0$ with an oscilloscope can potentially shed more light on audio performance than just deducing a nominal $\tau_c$ from $f_c$ (i.e., spectral analysis). This point was illustrated for audio cable characteristics in [177] and is shown here in Fig. 22: From panel (a), the 90%-to-10% fall time $\tau_{fall}$ of cable G is shorter than for cable S ($\tau_{fall} = 300$ ns); but G has almost double the 60-dB fall time ($\tau_{60}=1666$ ns) compared to the $\tau_{60} = 936$ ns for S (panel (b)) due to its non-ideal capacitive behavior. Furthermore, the response for S has cleanly disappeared (below this measurement's threshold) by 1.1 μs, whereas G still has observable residue at 2.4 μs. It is important to note that this type of distortion will not show up in a frequency-spectrum measurement: the measured gains and phases varied by less than ±0.03 dB and ±0.06 degrees respectively for both cables (see Fig. 5 of [177]).

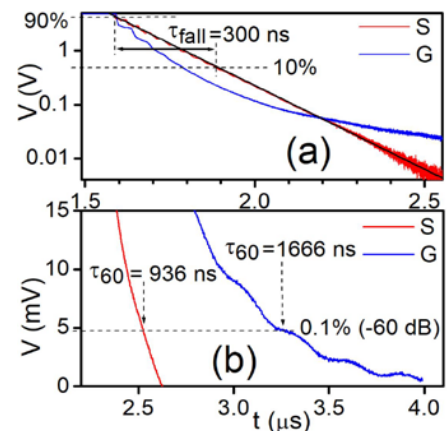

*Fig. 22. Decay of voltage after a downward step (at 1.59 μs) for two interconnect cables S and G [177]. (a) Extended voltage range. The ideal capacitive behavior of S produces an exponential decay (straight line). (b) Low-voltage-range measurement with extended time scale. Further details can be found in [177].*

Along the same lines, Fig. 21(c) shows digital-audio artifacts due to pre and post ringing for times approaching ~ 1 ms, tracing the signal to very low levels as one should. This extended-time low-level response should be taken into consideration, along with the FWHM, when designing the filter response.

### 3.11 IC and SC: Integration, categorization, and mapping

So far there has been an outward branching of information from the ANFs to various brain-stem stations (SOC, LL, etc.) which parallel process different basic tasks (ITD, ILD, edge detection, etc.). This information converges together in the IC, which has neurons specialized in how they respond to specific durations, temporal sequences, frequency combinations, etc. The sounds are differentiated by various characteristics and patterns such as waveform envelope, AM rate, FM rate, FM-sweep direction, and direction of motion [179] [180] [181] [182]. Responses to moving sources are dependent on the history of the stimulus and other inputs such as from the visual system. Some cells display a "novelty response", habituating after a few repetitions of the same stimulus, and responding again if parameters change [183].

In the MSO's ITD circuitry, delays were incorporated through differences in nerve-fiber length and synaptic delays. Those delays are relatively short (microseconds to milliseconds). IC encodes longer temporal features through inhibitory neurons in the delay lines and neurons with slower internal temporal responses, allowing processing of complex sequences of IPSPs and EPSPs lasting well beyond the duration of simple sounds. Modulation of IC neuronal processing characteristics over even longer periods (hours) takes place through descending cortical feedback, which is primarily excitatory but can provide inhibition through intermediate inhibitory interneurons [182] [184].

We saw earlier how frequency selectivity is sharpened in the cochlea by the CA system, and in the VCN through AND gating. Further sharpening takes place in the IC through suppression of the flanks of the tuning curves by inhibitory inputs, which create band-pass filters for frequency and level, as well as play a role in temporal processing [180] [182] [185]. While the IC may form a rudimentary map of distances between sound sources, it is in the *superior colliculus* (SC, a mainly visual processing center) that auditory information from the IC, along with visual and somatosensory[47] information, forms topographic maps based on source locations [186]. These maps are aligned between the senses[48] and the SC motor areas, to facilitate integration between the senses and initiate appropriate motor responses.

## 4 HIGHER BRAIN CENTERS AND MEMORY

All auditory information—mostly coming through the IC but some directly from brainstem nuclei—ascends through the MGB[49] (*medial geniculate body*) in the thalamus before entering the *auditory cortex* (AC). MGB continues and extends the IC's function, but holds a more bidirectional partnership with the AC in extracting and bridging together features identifying higher-order sound-element combinations (e.g., syllables and words in the case of speech [187] [188]).

The cerebral cortex (containing ~100 billion neurons with ~100–1000 trillion synaptic connections) represents the highest level of our nervous system and the outermost portion of the brain. It has a highly convoluted structure sculpted by *gyri* (ridges) and *sulci* (grooves), which compactify its ~1 m$^2$ area and reduce intracortical distances for faster communication. It is separated by *fissures* into hemispheres and lobes (principally frontal, temporal, parietal, and occipital). About ~90% of human cortex consists of 6 layered neocortex and ~10% of 3-4 layered allocortex. The hemispheres are connected by the *corpus callosum*, and each hemisphere responds mainly to the opposite-side ear because most ANF signals cross over to the contralateral side before reaching the cortex. Whereas subcortical stations are organized into somewhat rigid functional nuclei, the cortex is organized into more plastic fields or areas[50]. Unlike the relatively detailed cellular-level knowledge of brainstem circuitry (e.g., Fig. 16, Fig. 18, and Fig. 20), our cortical knowledge (especially for humans) mainly consists of which fields are active for various features of sounds. Generally, less is known about AC than its visual counterpart.

AC is located in the upper (superior) portion of the temporal lobe. It consists of a *primary* (core) field A1 (located in Heschl's gyrus, HG) surrounded by various *association* (belt and parabelt) areas that provide further processing and interpretation [188]. A1 and some of the other fields[51] maintain tonotopic arrangement that traces back to the cochlea. A1 neurons are also tuned by other characteristics such as level and spatial direction. Aspects such as timbre, pitch height, and pitch chroma are mapped in independent association areas [189] [190].

Pure-tone pitch may simply be represented through the tonotopic map in A1. But determination of complex-tone pitch is not understood; although, which brain areas activate for sounds with pitch salience or other specific attributes has been determined through mathematical decomposition of fMRI images of a variety of sounds and through intercranial recording with electrodes [191] [192]. Complex tones evoke the pitch of the "implied fundamental"—i.e., the periodicity of the waveform in 'time' or the spacing between harmonics in frequency ('places' on the BM)[52]. It is believed that some combination of these 'time' and 'place' mechanisms is operative in pitch determination, with a probable bias toward the former [193] [194] [195].

While subcortical levels, starting with the cochlea, already facilitate high frequency selectivity, further sharpening occurs in MGB and AC, where the tuning is also more robust (i.e., independent of sound level) compared to lower centers [196] [197]. It can thus be expected that temporal resolution will also be further refined in the cortex. A crucial task of the cortex is *auditory scene analysis*, whereby punctuation features such as temporal onset delineate individual auditory events. Research ranging from single neuronal measurements to the psychophysics of amplitude-transient detection and masking indicates that temporal-edge detection is encoded in cortical onset response [188] [198] [199] [200] [201]. Tones showing degradation at lower levels when mixed with noise are restored in the cortex, especially if the noise is temporally gated with the tone. Behavioral studies have shown that the temporal envelope, even with faulty spectral information, was sufficient for speech perception [202].

Because final feature detection of sounds takes place in the cortex, there can be significant differences in ability to notice sonic details that is independent of peripheral hearing performance. Thus elderly individuals missing one or two octaves of $f_{max}$ may be able to distinguish minute differences in fidelity that may be unnoticeable to young people with perfect audiograms[53]. The visual counterpart of this is *prosopagnosia* (face blindness) in which patients are unable to distinguish faces despite otherwise perfect vision. Conversely, a patient with cortical deafness can be unaware of sounds (i.e. not "hear" them) but still respond reflexively to sounds since lower brain stages (e.g., SC) have direct connections to motor functions.

---

[47] *Somatosensory* refers to sensations such as touch, pressure, vibration, movement, position, pain, and temperature, which originate in the skin or from points within the body such as joints or muscles.

[48] Vision plays an important role in calibrating auditory mapping during infancy [186].

[49] Visual information from the SC enters the cortex through the LGB (*lateral geniculate body*).

[50] The CN shows basically no plasticity, but intermediate stations (e.g., IC) have some ability for reorganization.

[51] For example in the cat, AAF and PAF (anterior and posterior auditory fields) maintain tonotopy whereas A2 (large ventral auditory field) does not [188].

[52] In the well-known *missing fundamental effect*, the fundamental frequency and some low harmonics can be removed without altering the pitch. E.g., the harmonic sequence 200, 300, 400 Hz…evokes the pitch of the 100 Hz highest common factor even though 100 Hz is absent.

The two hemispheres (and hence opposite ears) emphasize different sonic features. Some evidence suggests the left side as being more adept at processing fine temporal structure and the right side at spectral processing (see [203] and box 1 of [204]). And OAEs and AEPs indicate superior right ear function versus left as discussed earlier. Listening tests involving just one ear may want to take these factors into consideration.

Information received through the senses is held fleetingly in *sensory memory* (SM), which is termed *echoic memory* for sound (responsible for persistence of sound and backward masking[54]) and *iconic memory* for vision (leading to persistence of vision). Echoic SM persists for ~0.2 s [205] [206]. Paying attention to items in SM transfers them to *short-term memory* (STM). Because attentiveness varies greatly between individuals, so does the ability to discern minute differences in fidelity.

STM can hold about 4 items (formerly thought to be 7 ± 2 items [207]) for 15–30 s; however, the items can represent large *chunks* of organized information—e.g., letters grouped into words, or words into poems. The vocabulary of colorful adjectives (bright, visceral, etched, syrupy, etc.) used by audiophiles[55] [208] aids this chunking process, making it easier to remember and compare sounds. Manipulation and comparison of information takes place in *working memory* (WM). SM, STM, and WM are based on short-term changes in the neural network (synaptic connections). Because of the very limited capacity of STM and WM, detailed long-lasting impressions of sound quality can only be formed in LTM (*long-term memory*).

LTM, which is distributed throughout the cortex, results from more durable *long-term potentiation* (LTP) of the synaptic strength between neurons as well as, over longer times, reorganization of neural circuits themselves (addition and deletion of synapses). There is no known capacity limit for LTM. Successive experiences progressively refine this memory by fine tuning the connections through LTP and LTD (*long-term depression*). Thus the first glimpse of a new face may retain only the gender, forgetting other details almost immediately; but repeated exposures progressively improve facial recognition making it robust against changes in hairstyle, etc. Hence, forming a definitive and detailed opinion about an audio system's sonic performance is a long and slow process.

LTM consists of *declarative* (or *explicit*) *memory*—which one can recall and narrate—and *non-declarative* (or *implicit*) *memory*—involved in learning skills (e.g., riding a bicycle), conditioning (e.g., moving reflexively away from a threatening sound), and priming (automatic influence of one stimulus over another; e.g., response to the word "bone" after hearing "dog"). Declarative LTM results from transfer from STM/WM, facilitated by the hippocampus [209], and is further subdivided into two types: *episodic* (events/experiences that one can relive through recollection) and *semantic* (learning of facts). Recalling a sonic feature, say excessive bass, involves both the episodic memory of the sensation and emotion, and the semantic classification of the sound as "bottom-heavy". These components occupy different brain regions and selective damage can affect one and not the other [210]. There is an interplay between the two and semantic memory is strengthened when associated with episodic. Both fade progressively over time, and episodic details may be survived by only their semanticized description.

Three factors that strengthen formation and retrieval of LTM include: information organization, association with meaning, and imagery [211] [212] [213]. These aids are in fact used in subjective comparisons of sound quality through the adjectives such as "airy" or "bloated" [208] and through spectral grouping (e.g., "mid-bass" or "upper treble") and other types of groupings of impressions of sounds. Sleep plays an important role in consolidating and strengthening memories [214].

These collective factors explain why audiophiles spend weeks auditioning a component/system—the *extended multiple-pass* (EMP) listening protocol facilitates forming a consolidated opinion in durable LTM. It also explains why standard blind tests employing *short-segment comparison* (SSC) of back-to-back brief stimuli often fail [215] [216]—the vast RD drastically exceeds the "perceptual bandwidth" [217] [218] and extremely coarse STM that underlies SSC[56]. It also explains why training improves listening-test statistics (e.g., see [171]). Thus judging sound quality takes time!

---

[53] Musical training causes significant cortical changes. It enlarges the corpus callosum and shifts the emphasis from the right to the left hemisphere.

[54] A sound event can be masked from attention retroactively before its transfer from SM to STM.

[55] From a scientific standpoint, it will be good to confirm that these qualities can indeed be discerned psychoacoustically and eventually measured objectively.

[56] SSC may be adequate for simple tasks (e.g., judging JNDs) and simple stimuli (e.g., pure tones). However real-world audio components (e.g., cables) will sprinkle myriad alterations across the NEP due to multiple distortions (e.g., noise, reflection sequences, and non-ideal residual decays). Hence the need for EMP over the SSC protocol for audio-component comparisons.

## 5 CONCLUSIONS

### 5.1 General summary

This article reviews all stages of the human audition process and brings to light certain properties that are not widely recognized, some of which are highlighted below.

1. The standard pure-tone *audiometric range* of young healthy ears is from $f_{min}$ =16 Hz to $f_{max}$ =18 kHz. However, ultrasonic frequencies can be sensed through mechanisms such as heterodyning (non-linear mixing) and bone conduction. In their initial stages, noise-induced and age-related hearing loss destroy mostly OHCs, not IHCs, in which case the "lost" frequency channels may still be able to sense at a higher threshold.

2. The ear's cochlear output is represented by the *neural excitation pattern* of 30000 nerve fibers, originating from 3500 IHC channels, that differentiate frequency, level, phase and onset times. This NEP hosts an astronomical number of variations and *resolution of detail* RD. Even for elderly listeners whose $f_{max}$ is only 4 kHz, this RD is $>10^{32}$.

3. The cochlear output is influenced by the ear's non-linearity, various active-control mechanisms, and descending neural feedback from higher centers. For loud sounds, the *acoustic reflex* acts within ~10 ms to protectively tighten the ear drum and pull away the stapes. As IHC channels are steadily stimulated the *cochlear amplifier* enhances the frequency tuning, sensitivity, and dynamic range of the channels, in 3 stages occurring on ~15 μs, ~240 μs, and >1 ms time frames.

4. At the earliest stage of the onset of sound, before the cochlear amplifier and acoustic reflex have had time to act, the cochlear response is primed for broadband transient detection through the PVCN-VNLL pathway. Neurophysiological modeling and psychoacoustic experiments indicate that this *transient resolution* may be on the order of 1–10 μs. Since this TR arises from IHC action alone, hearing impaired listeners with mainly OHC loss may still have good TR and hence be able to well discern a musical instrument's attack transient.

5. At the cochlear level (IHC receptor potential), phase information exists only for f < 4 kHz. Above that what comes out is a voltage plateau (Fig. 4[b]) for which only an onset time (not phase) can be meaningfully defined. Monaural phase is largely ignored in timbre perception as enunciated by *Ohm's law of acoustics*. But relative onset timings between frequency components comprising the attack are timbre influential.

6. "Temporal resolution" is a broad umbrella term that includes a host of timing-information processes that make sense of musical sounds—ranging from the tempo and note lengths to slew rates of note attacks—as well as detecting odd features (unrelated to musical sounds and audio distortions) such as gaps in sinusoids. For clarity, the term *transient resolution* is being used for the discriminability of impulses and onsets (attacks).

### 5.2 Implications for audio

7. Audible-band *frequency response* and *linearity* (deviation from which is reflected in time-correlated distortions such as harmonic and intermodulation) may be of some value for discriminating entry-grade consumer audio equipment, but are relatively useless for high-end audio equipment, all of which is already sufficiently close to perfection in these respects (although less so for loudspeakers). At the standard of HEA, sonic differences are more likely to arise from various time-domain distortions (principally the *temporal smear* τ and the decay *cutoff time* $τ_c$) or extension of FR into the ultrasonic range. Hence circuit designs using (time-lagging) negative feedback to improve frequency response and linearity at the expense of time-domain performance can be expected to degrade sonic performance.

8. While ultrasound (i.e., f > $f_{max}$) may not be audible at moderate levels when played one pure frequency at a time, it can be audible at high levels or as part of a complex tone due to mechanisms such as heterodyning.

9. While time- and frequency-domain representations of a *signal* are perfectly transmutable through the Fourier transform/inverse-transform, this does not hold for a system's *response* (transfer function) except for an idealized linear and time-invariant system. The response of the ear and audio equipment depends on the structure, level, and history of the signal. Hence τ and $τ_c$ cannot be exactly deduced from the FR, and a spectrum analyzer using continuous sinusoidal signals cannot reveal the same information as an oscilloscope.

10. Based on earlier points 4 and 8, it can be roughly estimated that a HEA component may need τ ~ 1–10 μs and $τ_c$ ~ 10–100 μs to be sufficiently transparent—conditions that may be satisfied by some cables and (pre) amplifiers, but probably not by most source components nor speakers.

11. Claims that differences in upstream components (e.g., source or amplifier) can be heard even when the system is bottle-necked by a mediocre downstream component (e.g., speaker) shouldn't seem surprising—given that the NEP can resolve 1 part in $10^{40}$.

12. Although the auditory system seems to have capabilities that might be hard to match in measurements, a researcher has the luxury of endlessly averaging repetitive signals and employing range-splitting to enhance SNR and DR. This is illustrated in Fig. 10 (c) where averaging of ~100000 spectra over numerous days achieved noise floors below 0 dB SPL.

13. The SSC (*s*hort *s*egments *c*ompared back-to-back) listening protocol, may be adequate for simple tasks such as detecting level changes in a sinusoid. Real-world audio distortions sprinkle myriad tiny variations all over the NEP. This complex pattern of change cannot be handled by the extremely limited short-term memory. Blind listening tests for comparing subtle differences between HEA components requires an EMP (*extended multiple passes* of listening to complex music) protocol. Having a "palate cleansing" break (preferably ~1 minute or longer) between stimuli resets short-term memory and recruits the durable and infinitely more detailed long-term memory.

14. An individual's audiogram does not convey the full scope of their ability to discern sonic details. Noise-induced and age-related hearing loss raise thresholds for hearing certain frequencies without necessarily seriously compromising TR and RD. Well performing feature-detection circuity at the cortical level and a detailed long-term memory of live sound, etched through a lifetime of concerts, can make an elderly listener more adept at noticing differences in audio quality than a less experienced young listener.

15. A lot of the controversy surrounding high-end and high-resolution audio arises because most of the community is unaware of many basic and essential facts about human hearing. From the published literature, it appears that even some auditory-temporal-resolution research studies are unaware of the synchronous AND gating processes taking place in the octopus neurons of the PVCN and their incorporation as an attack-assessment step in pattern-recognition in the VNLL. It is hoped that the present work will bring wider awareness and appreciation of the complexities and intricacies of the human auditory system, so that future analyses of audio performance will be based on a better biological foundation.

# 6 ACKNOWLEDGMENTS

This work has benefitted from discussions with professionals in many fields (listed alphabetically): acoustics, audio engineering, biomedical engineering, communication sciences, musicology, neuroscience, otolaryngology, physics, physiology, and psychology. The following are gratefully acknowledged for their valuable feedback and interactions (alphabetically by last name): Meisam Arjmandi, Joe Azar, Reginald Bain, Martin Colloms, Charles L. Dean, Anjali R. Desai, Rutvik H. Desai, William M. Hartmann, Wilbert Van Meter Johnson, James M. Knight, Peter Lindenfeld, David Mott, Donata Oertel, Jan-Eric Persson, Matthew W. Rhoades, Thomas D. Rossing, Gabriel F. Saracila, Grigory Simin, Stacy D. Varner, Douglas H. Wedell, and Fan-Gang Zeng. I also thank the reviewers who made many helpful suggestions to improve the article.

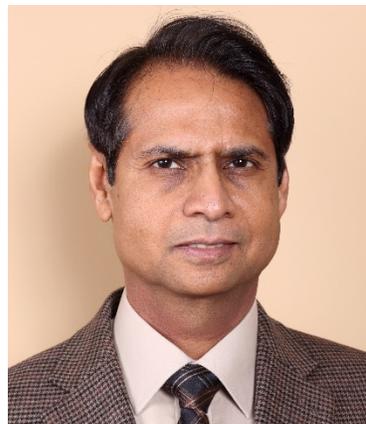

Milind N. Kunchur is a Governor's Distinguished Professor and Michael J. Mungo Distinguished Professor at the University of South Carolina in Columbia, U.S.A. He is a Fellow of the American Physical Society and has won a Carnegie Foundation U.S. Professors of the Year award. He was named a Governor's South Carolina Professor of the Year and has received the George B. Pegram Medal, Ralph E. Powe Award, Donald S. Russell Award, Martin-Marietta Award, Michael A. Hill Award, Michael J. Mungo Award, and held a National Research Council Senior Fellowship.